\documentclass[twocolumn,amsmath,amssymb,aps]{revtex4-2}
\usepackage{lineno,hyperref}
\usepackage{bm}
\usepackage{amsmath}
\usepackage{gensymb}
\usepackage{epstopdf}
\usepackage{booktabs}

\usepackage{graphicx}
\usepackage{ulem}
\usepackage{xcolor}
\usepackage{hyperref}
\begin{document}

\title{High spin-polarization in a disordered novel quaternary Heusler alloy FeMnVGa }


\author{Shuvankar Gupta$^1$}
\email{guptashuvankar5@gmail.com}
\author{Sudip Chakraborty$^1$}
\author{Vidha Bhasin$^2$}
\author{Santanu Pakhira$^3$}
\author{Shovan Dan$^4$}
\author{Celine Barreteau$^5$}
\author{Jean-Claude Crivello$^5$}
\author{S.N. Jha$^6$}
\author{Maxim Avdeev$^{7,8}$}
\author{Jean Marc Greneche$^9$}
\author{D. Bhattacharyya$^2$}
\author{Eric Alleno$^5$}
\author{Chandan Mazumdar$^1$}

\affiliation{$^1$Condensed Matter Physics Division, Saha Institute of Nuclear Physics,  A CI of Homi Bhabha National Institute, 1/AF, Bidhannagar, Kolkata 700064, India}
\affiliation{$^2$Atomic \& Molecular Physics Division, Bhabha Atomic Research Centre, Mumbai 400 094, India}
\affiliation{$^3$Ames Laboratory, Iowa State University, Ames, Iowa 50011, USA}
\affiliation{$^4$Department of Condensed Matter Physics and Materials Science, Tata Institute of Fundamental Research, Mumbai 400005, India}
\affiliation{$^5$Univ Paris Est Creteil, CNRS, ICMPE, UMR 7182, 2 rue Henri Dunant, 94320 Thiais, France}
\affiliation{$^6$Beamline Development and Application Section Physics Group, Bhabha Atomic Research Centre, Mumbai 400085, India}
\affiliation{$^7$Australian Nuclear Science and Technology Organisation, Locked Bag 2001, Kirrawee DC, NSW 2232, Australia}
\affiliation{$^8$School of Chemistry, The University of Sydney, Sydney, NSW 2006, Australia}
\affiliation{$^9$Institut des Mol\'{e}cules et Mat\'{e}riaux du Mans, IMMM, UMR CNRS 6283, Le Mans Universit\'{e}, Avenue Olivier Messiaen, Le Mans Cedex 9, 72085, France}

\date{\today}
\begin{abstract}
In this work, we report the successful synthesis of a Fe-based novel half-metallic quaternary Heusler alloy FeMnVGa and its structural, magnetic and transport properties probed through different experimental methods and theoretical techniques. Density functional theory (DFT) calculations performed on different types of structure reveal that the structure with Ga at 4\textit{a}, V at 4\textit{b}, Mn at 4\textit{c} and Fe at 4\textit{d} (space group: \textit{F$\bar{4}$3m}) possess minimum energy among all the ordered variants. \textit{Ab-initio} simulations in the most stable ordered structure show that the compound is a half-metallic ferromagnet (HMF) having a large spin-polarization (89.9 \%). Neutron diffraction reveals that the compound crystalizes in disordered Type-2 structure (space group: \textit{Fm$\bar{3}$m}) in which Ga occupies at 4\textit{a}, V 4\textit{b} and Fe/Mn occupy 4\textit{c}/4\textit{d} sites with 50:50 proportions. The structural disorder is further confirmed by X-ray diffraction (XRD), extended X-ray absorption fine structure (EXAFS),${^{57}}$Fe M\"{o}ssbauer spectrometry results and DFT calculations. Magnetisation studies suggest that the compound orders ferromagnetically below $T_{\rm C}$ $\sim$ 293 K and the saturation magnetization follows the Slater-Pauling rule. M\"{o}ssbauer spectrometry, along with neutron diffraction suggest that Mn is the major contributor to the total magnetism in the compound, consistent with the theoretical calculations, which also indicates that spin-polarization remains high (81.3 \%), even in the presence of such large atomic disorder. The robustness of the HMF property in presence of disorder is a quite unique characteristic over other reported HMF in literature and makes this compound quite promising for spintronics applications.
\end{abstract}
\maketitle

\section{\label{sec:Introduction}Introduction}
 In the realm of spintronics, half-metallic ferromagnetic (HMF) materials are one of the most promising systems that can potentially generate 100 \% spin-polarised current due to their unusual band structure~\cite{felser2007spintronics,wolf2001spintronics}. In such systems, one spin sub-band acts like a metal whereas the other sub-band behaves like a semiconductor~\cite{de1983new,park1998direct,jourdan2014direct}. Heusler alloys have generated a lot of interest because of their tunability among the several types of HMF materials that have been reported~\cite{bombor2013half, graf2011simple}.

The stoichiometric representation of the Heusler alloy is $X_2YZ$, where $X$,$Y$ are transition elements and $Z$ is \textit{sp}-group element~\cite{graf2011simple}. Recently, it was found that the quarternary variant of Heusler alloys, $XX^{\prime}YZ$, have a great potential to simultaneously exhibit the spin-gapless semiconducting (SGS)~\cite{ouardi2013realization,bainsla2015spin,bainsla2015origin} and HMF properties, which have enormously intrigued the interest of the scientific community~\cite{bainsla2016equiatomic,alijani2011electronic,alijani2011quaternary,venkateswara2019coexistence}.

 SGS's are a type of HMF in which the metallic sub-band only touches the Fermi level instead of complete overlapping, while the other sub-band remains as a semiconductor~\cite{wang2008proposal,ouardi2013realization}. Incidentally, most members of this sub-class of Heusler alloys are found to be Co-based. Although the theoretical predictions have extended to many other Heusler alloys beyond the Co-based analogues ~\cite{kundu2017new,gao2019high}, only a few have been experimentally realized so far, exhibiting HMF and SGS properties~\cite{bainsla2015spin,bainsla2015origin}. It is primarily because, most of those theoretically proposed compositions were often found to be experimentally untenable to crystallize. Even those which could be formed in single phase, the majority was found to consist of multiple elements belonging to the same period of the periodic table which is highly conducive to the cross-site occupation, resulting in significant structural disorder~\cite{graf2011simple}.
 For example, MnCrVAl has been theoretically predicted to be a SGS when formed in a perfectly ordered crystal structure~\cite{ozdougan2013slater,gao2019high}. However, the compound were found to form with considerable atomic disorder and without the expected SGS characteristics. The absence of SGS behaviour in this material is attributed to the site-disorder present in the actual compound~\cite{herran2017atomic,kharel2017effect}. Similar circumstance also occurs for the compound NiFeMnSn which was theoretically predicted to be a HMF, but in the presence of large structural disorder, the compound becomes metallic, instead of HMF~\cite{mukadam2016quantification}. It is therefore quite logical and sensible that one either has to prepare the compound as defect-free or try to identify the compounds where the SGS and HMF properties remain robust despite the presence of site-disorder(s).

 In this work, we have extended our search for new quarternary Heusler alloys beyond the Co-based systems and report the synthesis and physical properties of a new compound, FeMnVGa. We have determined the structural disorder using different local and bulk probes \textit{viz}. M\"{o}ssbauer spectrometry, Extended X-ray Absorption Fine Structure (EXAFS) and neutron diffraction. First principle calculations in both ordered and disordered phases reveal that the spin-polarization of the material is robust with respect to structural disorder. The magnetic and transport properties have also been investigated using different experimental tools.

\section{\label{sec:Methods}Methods}

\subsection{Experimental}
The polycrystalline FeMnVGa was prepared by a arc melting procedure taking suitable high purity ($>$99.9 \%) component elements. To obtain better homogeneity, the sample was melted 5-6 times under argon environment and flipped after each melt. During the melting process, an extra 2\% Mn was added to compensate for its evaporation. For structural characterization, both neutron and X-ray diffraction methods have been used. Powder neutron diffraction ($\lambda$ = 2.43 {\AA}) data were collected at 400 and 3 K at ECHIDNA beamline in ANSTO, Australia~\cite{avdeev2018echidna}. Cu-K$\alpha$ radiation on X-ray diffractometer (Model: TTRAX-III, M/s Rigaku Corp., Japan) was used to obtain powder X-ray diffraction (XRD) pattern at room temperature. The sample's single-phase nature and crystal structure were determined by performing Rietveld refinement of neutron diffraction as well as X-ray diffraction data using the FULLPROF software programme~\cite{rodriguez1993recent}.

EXAFS measurements of the FeMnVGa samples were carried out at the Energy Scanning EXAFS beamline (BL-9) at Indus-2 Synchrotron source (2.5 GeV, 300 mA) at the Raja Ramanna Centre for Advanced Technology (RRCAT), Indore, India. The beam line operates in the photon energy range of 4-25 keV. In this beamline a Rh/Pt coated meridional cylindrical mirror is used for vertical collimation of the beam coming from the storage ring. The collimated beam is monochromatized by a Si (111) (2d = 6.2709 {\AA}) based double crystal monochromator (DCM). The second crystal of the DCM is a sagittal cylindrical crystal, which is used for horizontal focusing of the beam while another Rh/Pt coated bendable post mirror facing down is used for vertical focusing of the beam at the sample position. Rejection of higher harmonics content in the incident X-ray beam is done by detuning the second crystal of the DCM. In the present case, the measurements have been carried out in fluorescence mode where the sample is placed at 45 degrees to the incident X-ray beam, and a fluorescence detector is placed at right angle to the incident X-ray beam to collect the signal. One ionization chamber detector is placed prior to the sample to measure the incident flux (I$_0$) and florescence detector measures the fluorescence intensity (I$_f$). In this case, the X-ray absorption coefficient ($\mu$) was obtained using the relation:  I$_T$ = I$_0$e$^{-{\mu}x}$, where x is the thickness of the absorber, and the full absorption spectrum was obtained as a function of energy by scanning the monochromator over the specified range.

Magnetic measurements were carried out in a commercial MPMS3 (Quantum Design Inc., USA) in the temperature range of 3$-$380 K and magnetic fields of up to $\pm$ 70 kOe. ${^{57}}$Fe transmission M\"{o}ssbauer spectrometry was used to investigate the atomic scale local environment and nuclear hyperfine structure of Fe. An electromagnetic transducer with a triangular velocity shape and a ${^{57}}$Co source diffused into a Rh matrix and a bath cryostat were utilized to acquire spectra at 300 K and 77 K. The samples are made up of a thin coating of powder with a Fe content of around 5 mg Fe/cm$^2$. Using the in-house software `MOSFIT', the hyperfine structures were simulated using a least square fitting approach employing quadrupolar doublets formed of Lorentzian lines. The isomer shift values were compared to those of ${\alpha}$--Fe at 300 K, and the velocity was regulated with an ${\alpha}$--Fe foil standard. Four-probe resistivity measurements were carried out in a commercial Physical Property Measurement System (Quantum Design Inc., USA).

\subsection{Computational}

The density functional theory (DFT) is used to calculate the stability and ground-state properties of the material. The projector augmented wave (PAW) approach~\cite{blochl1994projector} included in the Vienna \textit{ab-initio} simulation package (VASP) was used to carry out DFT calculations~\cite{kresse1993ab, kresse1994norm}. The exchange correlation was described by the generalized gradient approximation modified by Perdew, Burke and Ernzerhof (GGA-PBE)~\cite{perdew1996generalized}. All calculations included an energy band up to a cutoff of E = 600 eV. The tetahedron approach with Blöchl correction~\cite{blochl1994improved} was used after performing volume and ionic (for disordered structure) relaxation procedures. All calculations were done taking spin-polarization in consideration. Unit cells based on the notion of special quasirandom structure (SQS)~\cite{zunger1990special} were created to model the numerous possible disorder schemes in FeMnVGa in order to model statistical chemical disorder. The cluster expansion formalism for multicomponent and multisublattice systems~\cite{sanchez1984generalized}, as implemented in the Monte-Carlo (MCSQS) algorithm provided in the Alloy-Theoretic Automated Toolkit (ATAT)~\cite{van2009multicomponent,van2013efficient}, was utilised to build the SQS. Subsequently, DFT calculations were performed to see how trustworthy the DFT results are and to test the quality of the SQS. Aside from the calculations with a distinct order of interactions, the root mean square (rms) error was utilised as another quality criterion. For all clusters \textit{k}, the rms error defines the departure of the SQS ($\Pi^{k}_{SQS}$) correlation function from the correlation function of a totally random structure ($\Pi^{k}_{md}$).
\begin{equation}
\ rms = \sqrt{\sum_{k}(\Pi^{k}_{SQS}-\Pi^{k}_{md})^2}
\label{eq1}
\end{equation}

To develop the disordered structure, several tests were performed to optimize the clusters type and their numbers (Ga at 4\textit{a} (0,0,0), V at 4\textit{b} (0.5,0.5,0.5) and Fe=0.5/Mn=0.5 at 4\textit{c} (0.25,0.25,0.25) and Fe=0.5/Mn=0.5 at 4\textit{d} (0.75,0.75,0.75) from the LiMgPdSn-type crystal structure (Space group: \textit{F$\bar{4}$3m}). Finally, to acquire reliable results, 7 pairs, 5 triplets, and 11 quadruplets interactions were considered. After several convergence tests, the SQS cell with 28 atoms (for comparison, details of SQS with many different number of atoms in the supercell are provided in Sec. IA of the Supplementary Material) and a 50:50 mixing on 4$c$ and 4$d$ sites, was considered to simulate the disordered structure.

\section{Results and Discussion}

\subsection{\label{sec:DOS_Ordered}Electronic structure calculations -- Ordered structure}

\begin{figure}[ht]
\centerline{\includegraphics[width=.48\textwidth,height=4cm,keepaspectratio]{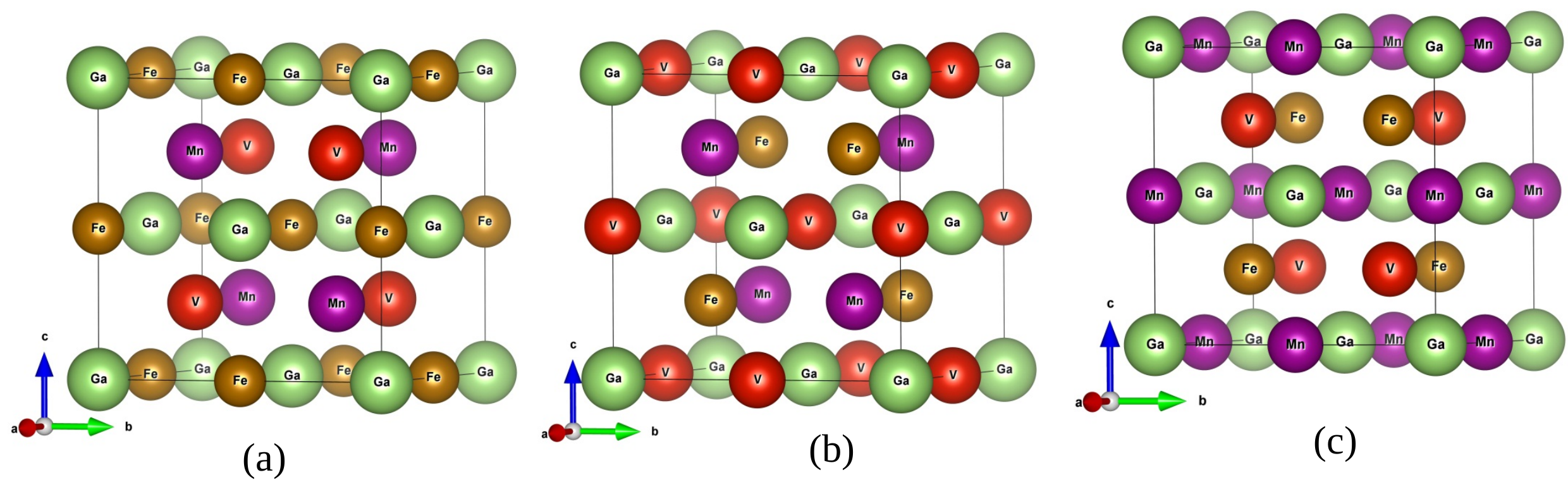}}
{\caption{Unit cell representation of (a) Type-1 (b) Type-2 and (c) Type-3 ordered structure as described in Table~\ref{Enthalpy}. Fe, Mn, V and Ga atoms are represented by yellow, magenta, red and green colors.}\label{All_Structure}}
\end{figure}

\begin{figure*}[t]
\centerline{\includegraphics[width=.96\textwidth]{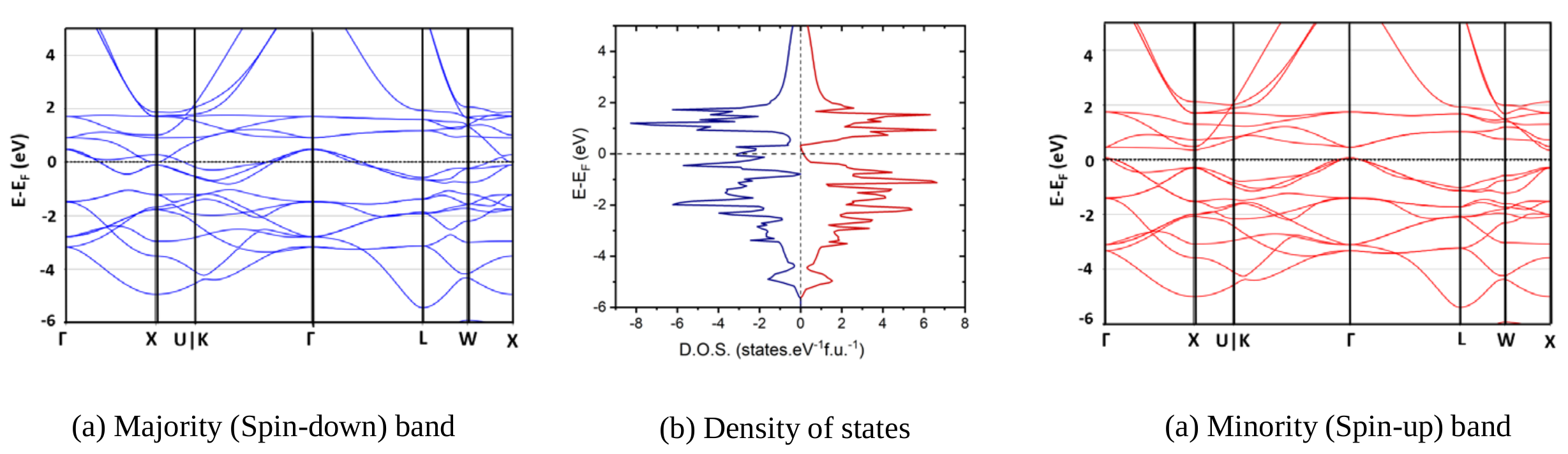}}
{\caption{Spin-polarized band structure and density of states of FeMnVGa in ordered Type-2 structure: (a) majority (spin-down) band (b) density of states, (c) minority (spin-up) band. The energy axis zero point has been set at the Fermi level, and the spin-up (minority) and spin-down (majority) electrons are represented by positive and negative values of the DOS, respectively.}\label{Ordered_DOS}}
\end{figure*}

DFT studies on FeMnVGa in LiMgPdSn-type structures (Space group: \textit{F$\bar{4}$3m}) were first done in order to optimise the crystal structure and identify the most stable configuration. In a quaternary Heusler alloy $XX^{\prime}YZ$, if the $Z$ atoms are considered at position 4$a$ (0,0,0),  the remaining three atoms $X$, $X'$ and $Y$ might be put in three alternative \textit{fcc} sublattices, namely 4$b$ (0.5,0.5,0.5),  4$c$ (0.25,0.25,0.25) and 4$d$ (0.75,0.75,0.75). Since the permutation of the atoms in the 4$c$ and 4$d$ positions leads to energetically invariant configurations, only three independent structures are viable, out of a total of six potential combinations. Fig.~\ref{All_Structure} depicts primitive representations of the three type of structures described in Table~\ref{Enthalpy}. In our calculations, we took into account these three combinations, and the results are presented in Table~\ref{Enthalpy}.\\

\begin{table}[]
\caption{Calculated enthalpy of formation $\Delta_f{H}$ for each ordered type of FeMnVGa, and one disordered case.}
\begin{tabular}{cccccc}
\hline \hline
 & 4\textit{a} & 4\textit{c} & 4\textit{b} & 4\textit{d} & $\Delta_f{H}$ (kJ/mol)\\ \hline
Type 1   & Ga & Mn & Fe & V  & -13.41                            \\
Type 2   & Ga & Mn  & V & Fe & -27.21                           \\
Type 3   & Ga & V & Mn  & Fe & -7.38 \\
disordered & Ga & Fe:Mn & V  & Fe:Mn & -28.58  \\ \hline \hline

\end{tabular}
\label{Enthalpy}
\end{table}

 According to our calculations, Type-2 structure is the most stable arrangement. Since the shortest atomic distance is along [111] direction, the three structure types should be first compared along this direction. Type-2 in the only type that prevents Fe and Mn atoms being nearest neighbors and promotes V and Ga atoms as their nearest neighbors. Based on this result and on a recent Bader analysis on the compound Fe$_2$VAl~\cite{diack2022influence}, it can be surmised that V (the most electropositive atom) and Ga atoms play the role of “cations” whereas Fe and Mn play the role of “anions” in FeMnVGa. The Type-2 structure, with its Ga – Mn – V – Fe sequence along [111], would be the only sequence alternating “cations” and “anions”. Total magnetic moment of full and quaternary Heulser alloy can be determined using Slater-Pauling rule given by m = (N$_V$-24) $\mu_{\rm B}$/f.u., where m is the total magnetic moment and N$_V$ is the total valence electron count (VEC)~\cite{galanakis2002slater}. Since VEC = 23 for FeMnVGa, following the convention provided by the Slater-Pauling rule, the total magnetic moment should be equal to -1 $\mu_{\rm B}$/f.u. It is crucial to note that these compounds with VEC $<$ 24 have negative total spin moments and that the gap is situated at the spin-up band as a result of the S-P rule~\cite{galanakis2007doping}. Additionally, as contrary to the other Heusler alloys~\cite{galanakis2002slater}, the spin-up electrons corresponds to the minority-spin electrons while the spin-down electrons match the majority electrons~\cite{galanakis2007doping}. The total magnetic moment is calculated to be -0.93\,$\mu_{\rm B}$/f.u., which is quite close to -1$\mu_{\rm B}$/f.u, as predicted by the Slater-Pauling (S-P) rule. The element-specific moments  have also been calculated for the most stable configuration (Type-2), with Fe = -0.56\,$\mu_{\rm B}$/f.u., Mn = -0.92\,$\mu_{\rm B}$/f.u., V = 0.53\,$\mu_{\rm B}$/f.u. and Ga = 0.02\,$\mu_{\rm B}$/f.u., resulting in ferrimagnetic coupling between z = 0 and z = 1/2 layers along c-direction. The predicted spin-polarized band structure and density of states (DOS) of the most energetically favourable configuration (Type-2 ordered structure) are shown in Fig.~\ref{Ordered_DOS}. The minority (spin-up) band exhibits an indirect band-gap at Fermi level (\textit{$E_{\rm F}$}), while the majority (spin-down) band is metallic. The current calculations show that FeMnVGa in the Type-2 structure has a very high polarisation $P=\frac{\rm{DOS}^\uparrow (E_{\rm F})- \rm{DOS}^\downarrow (E_{\rm F})}{\rm{DOS}^\uparrow (E_{\rm F})+ \rm{DOS}^\downarrow (E_{\rm F})}$ = 89.9\%, indicating that it is almost a half-metallic ferromagnet.

\subsection{\label{sec:Neutron}Neutron diffraction}
\begin{figure}[h]
\centerline{\includegraphics[width=.48\textwidth]{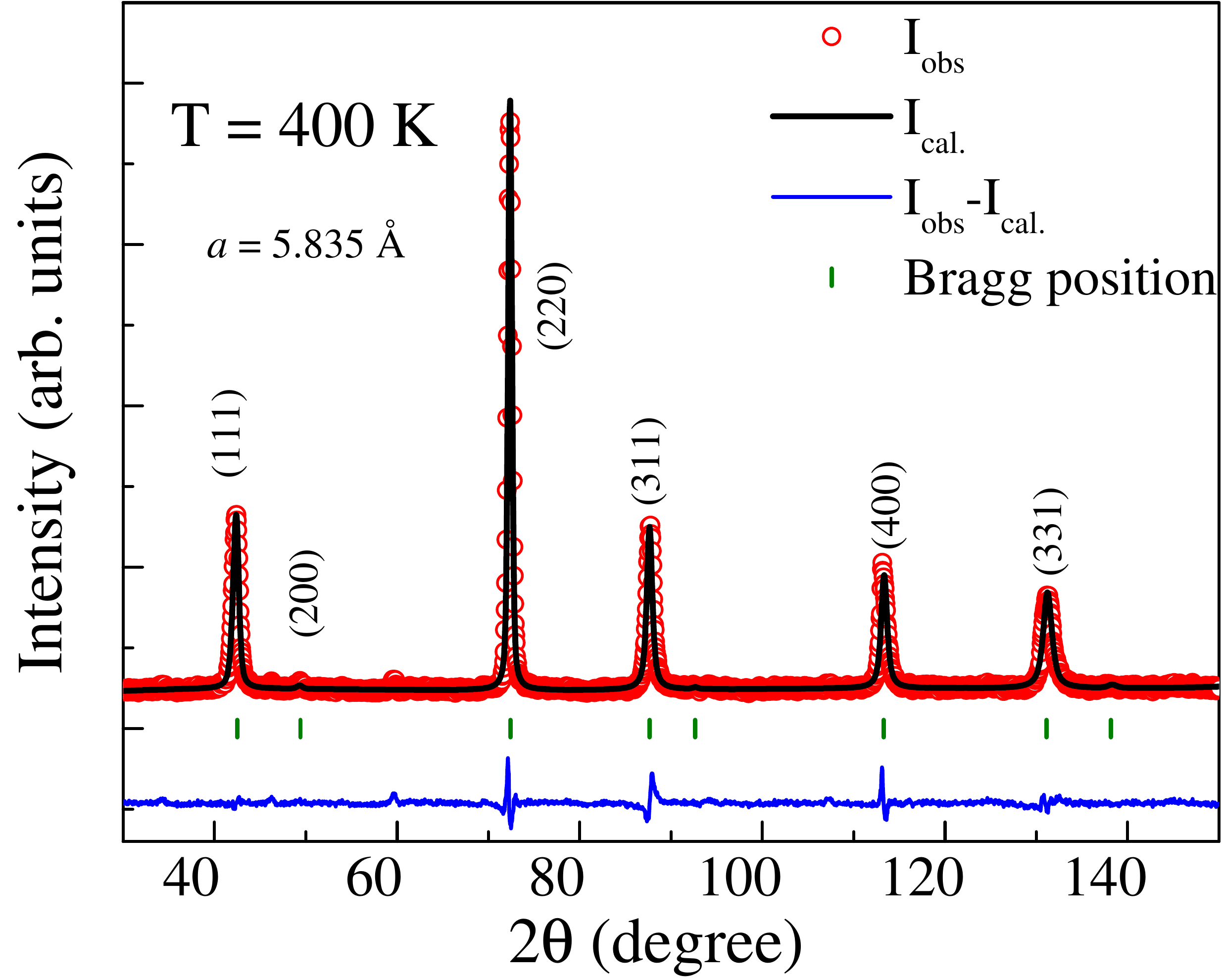}}
{\caption{Rietveld refinement of the Neutron diffraction pattern of FeMnVGa taken at 400 K.}\label{NPD_Fig}}
\end{figure}

\begin{figure}[ht]
\centerline{\includegraphics[width=.48\textwidth]{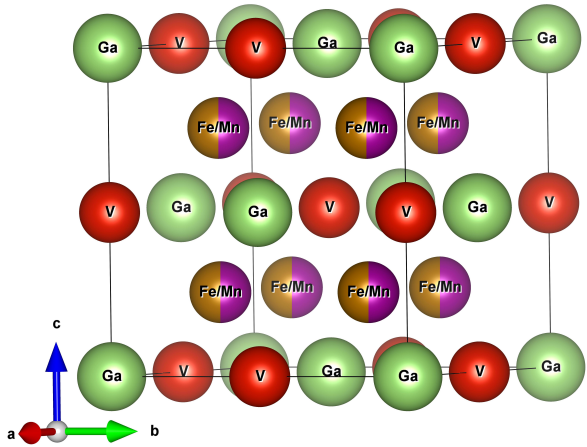}}
{\caption{Atom distribution of FeMnVGa in disordered crystal structure (L2$_1$-type, space group: \textit{Fm$\bar{3}$m}).}\label{Disorder_Structure}}
\end{figure}

In the studied compound FeMnVGa, all the constituent atoms (Fe, Mn, V and Ga) belongs from the same period of the periodic table. For such cases in Heusler alloy, X-ray diffraction often gives delusive results about the extent of atomic disorder within the crystal structure due to the close-by X-ray scattering factors of the constituent elements. Because of its capability to distinguish nearby atoms in the periodic table whose coherent scattering amplitudes in neutron diffraction process are dramatically different, neutron diffraction is generally more sensitive to site-disorder than X-ray diffraction technique~\cite{mukadam2016quantification,samanta2018reentrant,raphael2002presence}. The neutron diffraction pattern of FeMnVGa taken at 400 K (T$>$T$_{\rm C}$) is presented in Fig.~\ref{NPD_Fig}. It has already been mentioned before that one of the primary prerequisites for attaining high spin-polarization in Heusler alloys is to have a highly ordered structure since the structural disorder generally acts as a strong deterrent~\cite{miura2004atomic}. A2 and B2-type disorders are known to be the most frequently observed disorders in Heusler alloys. In A2-type disorder, all the constituent atoms ($X$, $X'$, $Y$ and $Z$) randomly mix with each other, whereas for B2-type disorder, the $Y$ $\&$ $Z$  and $X$ $\&$ $X'$ atoms randomly mix with each other in the 4\textit{a} $\&$ 4\textit{b}  and 4\textit{c} $\&$ 4\textit{d} sites, respectively~\cite{bainsla2016equiatomic,graf2011simple}. In general, the observation of (111) and (200) superlattice peaks signifies the ordered formation of the crystal. In an A2-type disorder, both superlattice reflections are missing, whereas only the (200) peak is present in a B2-type disorder~\cite{webster1973magnetic,graf2011simple}. Structural optimization calculations discussed above have proposed that Type-2 ordered structure having Ga-4\textit{a}, V-4\textit{b}, Mn-4\textit{c} and Fe-4\textit{d} possess minimum energy among all the probable ordered structural arrangements. However, neutron diffraction data taken at 400 K could not be explained with an ordered Type-2 arrangement. Interestingly, in the neutron diffraction data taken in the paramagnetic region (400 K), the (111) peak is clearly visible, ruling the possibility of both the A2 and B2-type disorders. It can thus be concluded that, it is neither intermixing of all the constituent atoms ($X$, $X'$, $Y$ and $Z$) in each sites nor the intermixing of $Y$ $\&$ $Z$  and $X$ $\&$ $X'$ atoms, in 4\textit{a} $\&$ 4\textit{b}  and 4\textit{c} $\&$ 4\textit{d} sites, respectively. However, the crystal structure can not be identified as ordered Type-2 structure as we find the (200) Bragg peak is diffused in the neutron diffraction data indicating the possible presence of another kind of atomic disorder. We have attempted to carry out the Rietveld refinement on the 400 K data with a disordered structure in which Ga is at 4\textit{a}, V at 4\textit{b} and Fe/Mn with 50:50 proportions at 4\textit{c} and 4\textit{d} sites, respectively. Such disordered structure presented in the Fig.~\ref{Disorder_Structure} perfectly fits the experimental data in the paramagnetic region (R$_{f}$ = 3.83, R$_{wp}$ = 5.12) (Fitting parameters for different disorder structure are provided in Supplementary material). Moreover, neutron diffraction also indicates 5\% of anti-site disorder between Ga and V sites. It is very fascinating to note that the disorder between Fe/Mn enhances the symmetry of the crystal structure and the disordered structure becomes L2$_1$-type (space group: \textit{Fm$\bar{3}$m}, no. 225) which is the ordered structure of ternary Heusler alloy.

\subsection{\label{sec:XRD}X-ray diffraction}

\begin{figure}[h]
\centerline{\includegraphics[width=.48\textwidth]{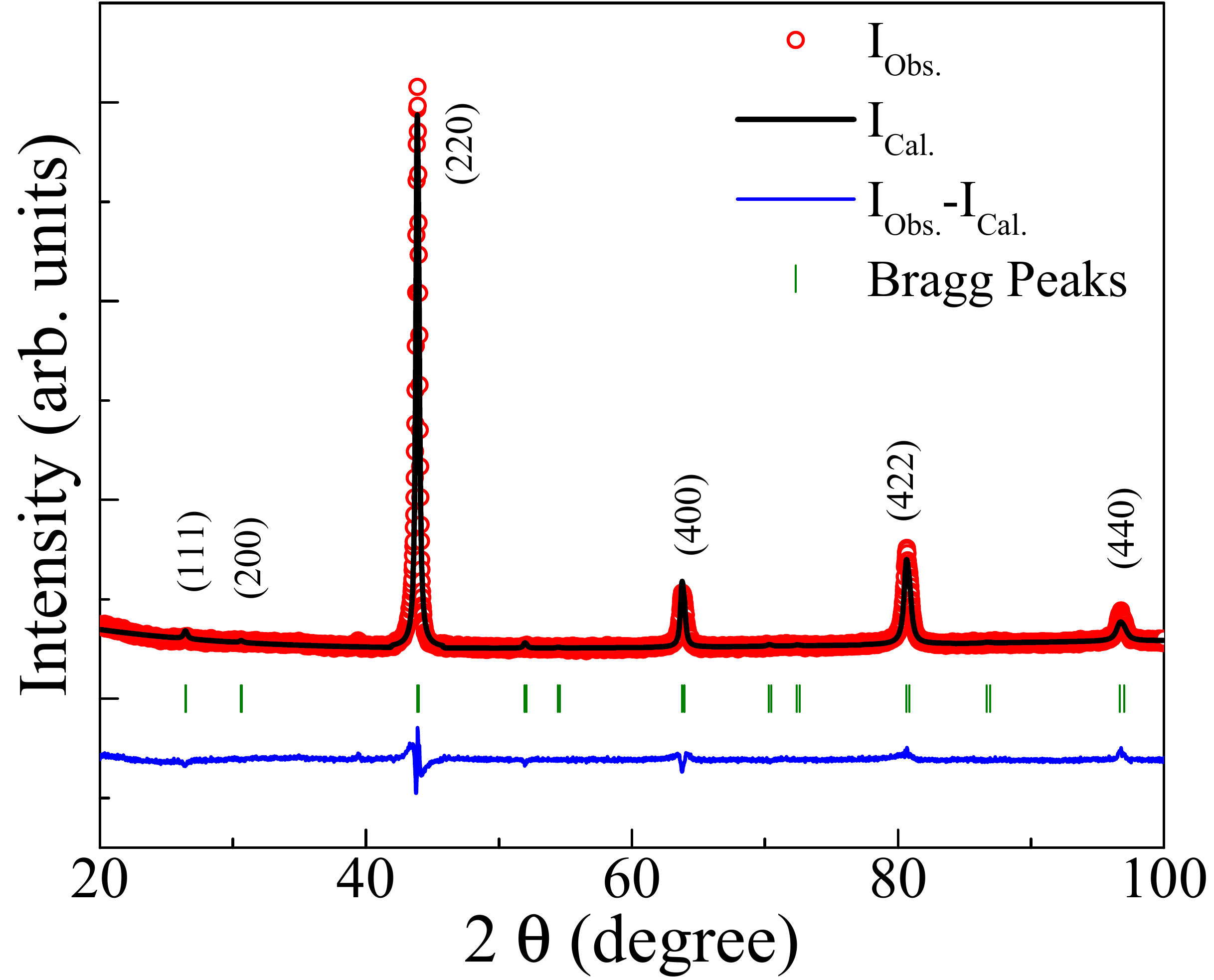}}
{\caption{Rietveld refinement of the powder XRD pattern of FeMnVGa disordered structure measured at room temperature. Miller indices of the Bragg peaks are presented in vertical first bracket.}\label{XRD_Fig}}
\end{figure}

To determine the lattice parameter more accurately (at room temperature), we have probed the system with X-rays, using a shorter wavelength compared to neutron. Due to Fe, Mn, V and Ga displaying very similar atomic X-ray scattering factors, an independent structural analysis of X-ray diffraction (XRD) data, not considering the structural model from neutron diffraction extracted above, could not be carried out. Any simulated X-ray patterns lead to the (111) and (200) lines being very weak or absent, making indistinguishable ordered or disordered structures~\cite{venkateswara2015electronic}. In such cases, one had no other option but to rely on the structural model extracted from the neutron diffraction data analysis. Fig.~\ref{XRD_Fig} represents the powder XRD data taken at room temperature and its Rietveld refinement, assuming the disordered structure (presented in Fig.~\ref{Disorder_Structure}) obtained from the neutron diffraction measurements. The calculated XRD pattern matches very well the experimental data (R$_{f}$ = 2.59, R$_{wp}$ = 3.61), showing that both diffraction experiments fully agree. The fitted lattice parameter is found to be a = 5.829 {\AA}. It may however be pointed out here that the neutron diffraction and XRD patterns analysis provides only the macroscopic information on the atomic disorders. To investigate the nature of disorders in the local atomic environment, we have utilized the extended X-ray absorption fine structure (EXAFS)~\cite{balke2007structural,bainsla2015local,ravel2002exafs} and ${^{57}}$Fe M\"{o}ssbauer measurements

\subsection{\label{sec:EXAFS}EXAFS}

\begin{figure}{}
\begin{minipage}{0.49\textwidth}
\centering
{\includegraphics[width=0.98\textwidth]{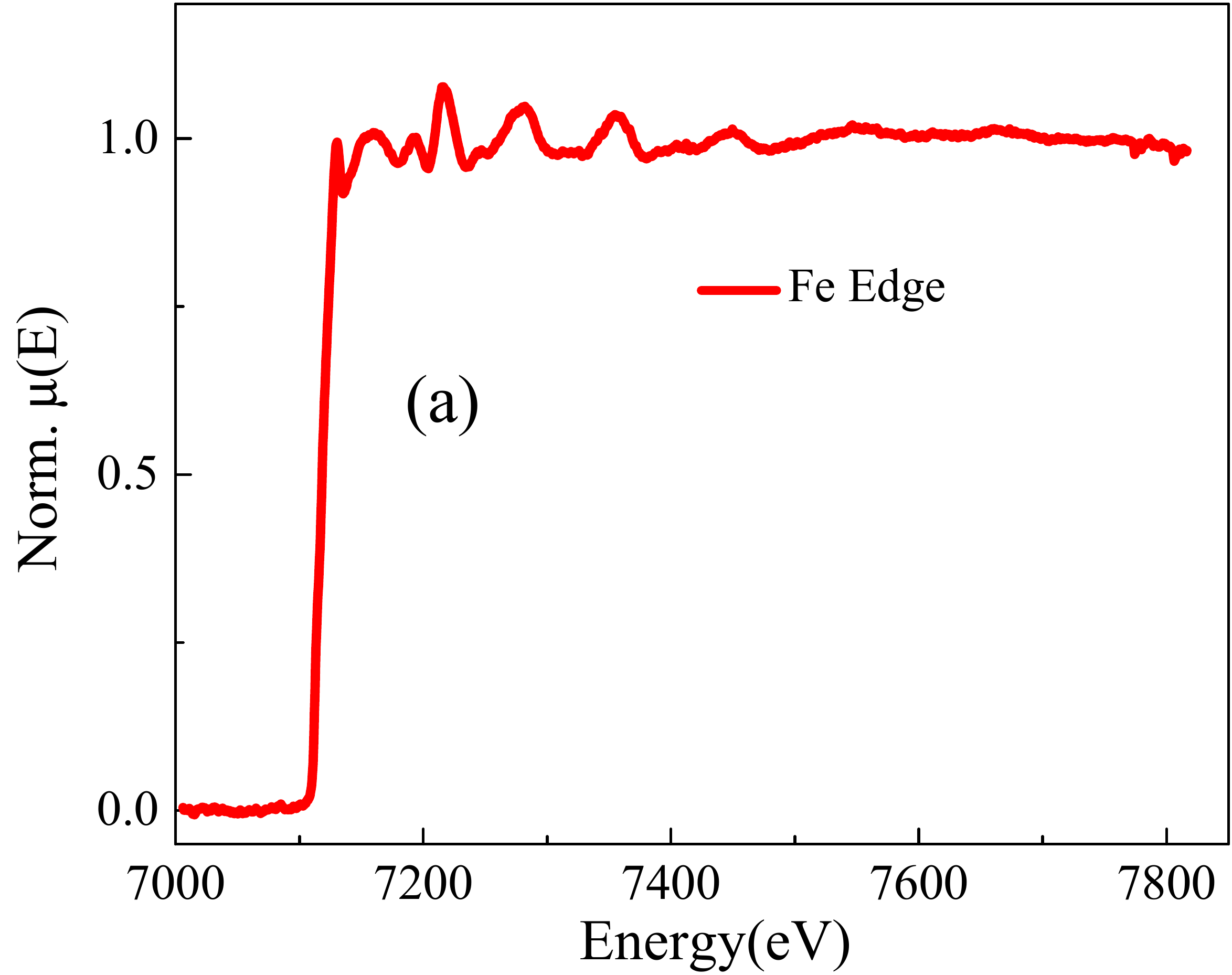}}\\
{\includegraphics[width=0.98\textwidth]{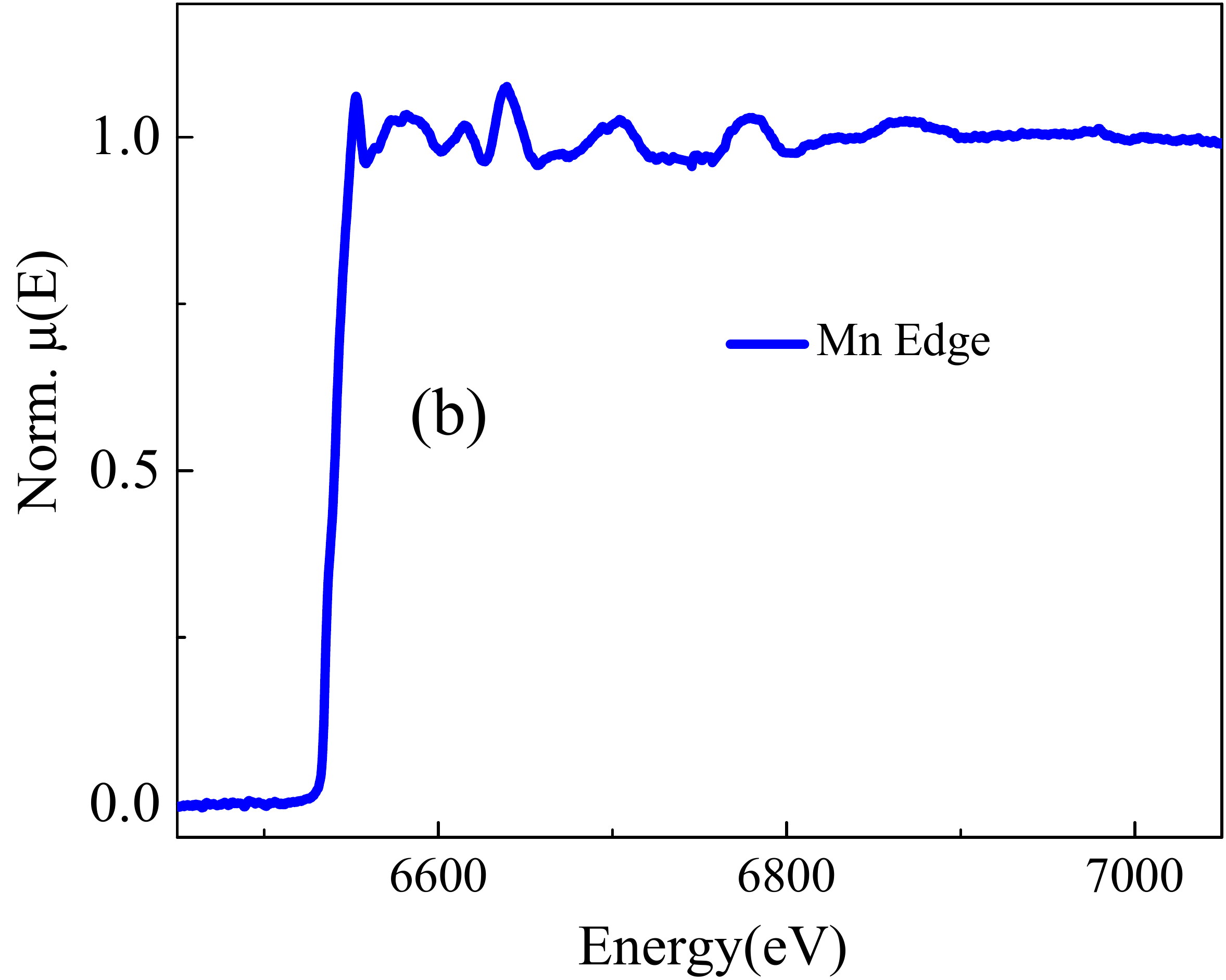}}\\
{\includegraphics[width=0.98\textwidth]{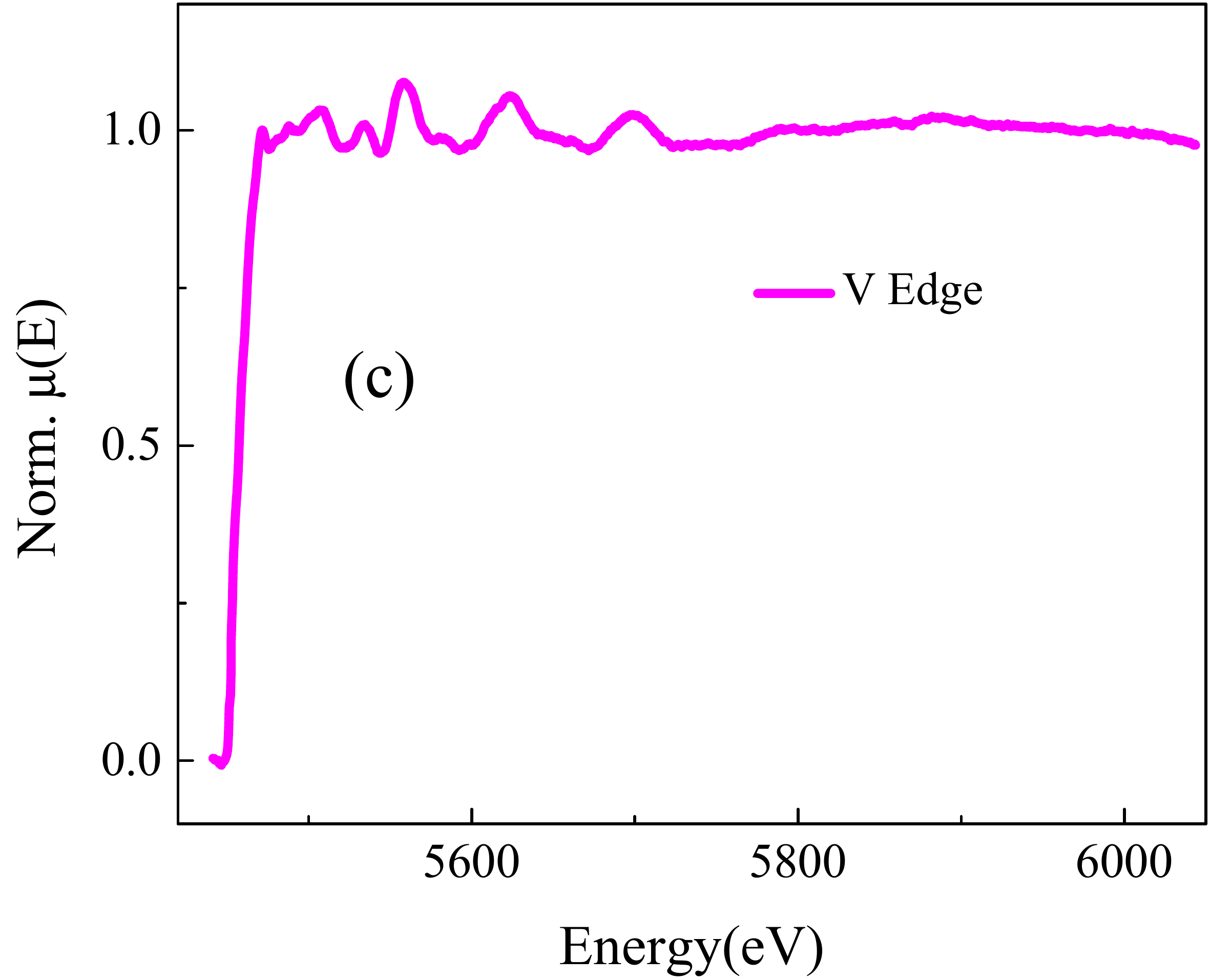}}
{\caption{Normalized EXAFS spectra of FeMnVGa taken at  (a) Fe edge (b) Mn edge and (c) V edge.}\label{EXAFS_Data}}
\end{minipage}
\end{figure}

Unlike XRD, extended X-ray absorption ﬁne structure (EXAFS) technique focuses on the atomic environment around the selected atoms. We conducted EXAFS measurements on FeMnVGa sample at the Fe, Mn, and V edges to investigate the structural disorder in a more restricted and local scale.

\begin{figure}[h]
\centerline{\includegraphics[width=.48\textwidth]{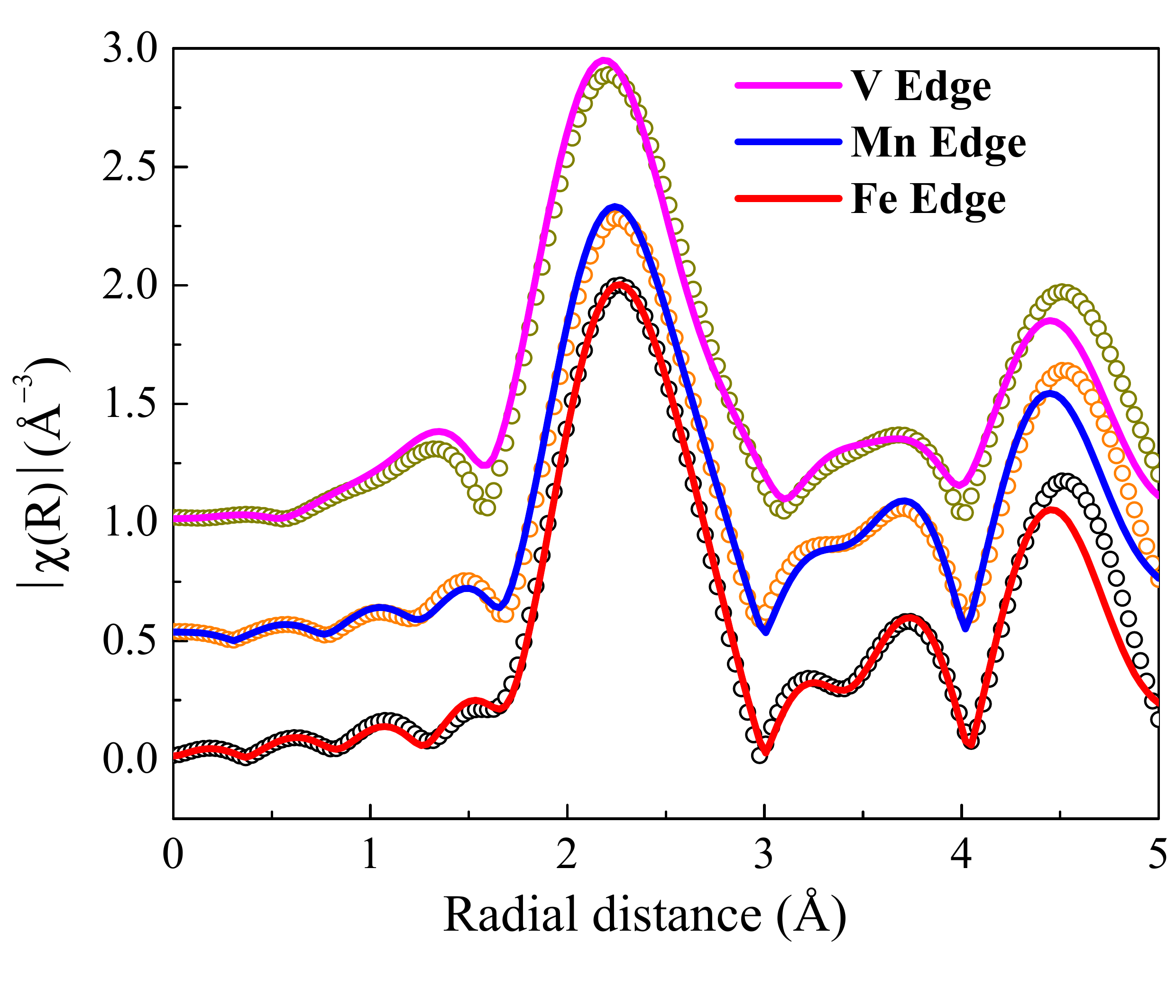}}
{\caption{Fourier transformed EXAFS spectra of FeMnVGa taken at Fe, Mn and V edges.}\label{EXAFS_FiT_Fig}}
\end{figure}

Fig.~\ref{EXAFS_Data} shows the normalised EXAFS ($\mu$(E) vs \textit{E}) spectra of the FeMnVGa sample measured at Fe, Mn, and V edges. Analysis of the EXAFS data have been processed following the standard procedure ~\cite{koningsberger1987x}.  In brief, to obtain quantitative information about the local structure, the absorption spectra ($\mu$(E) vs \textit{E}) has been converted to absorption function $\chi$(E) defined as follows:

\begin{equation}
\chi(E) = \frac{\mu(E)-\mu_{0}(E)} {\Delta\mu_{0}(E_{0})}
\label{eq1}
\end{equation}
\noindent
where \textit{E$_0$} is the absorption edge energy, $\mu$$_0$(\textit{E$_0$}) is the bare atom background and $\Delta$$\mu$$_{0}$(\textit{E$_{0}$}) is the absorption edge step in $\mu$(\textit{E}) value. The energy dependent absorption coefficient $\chi$(\textit{E}) is then converted to the wave number dependent absorption coefficient using the relation:

\begin{equation}
K=\sqrt {\frac {2m(E-E_{0})}{\hbar^{2}}}
\label{eq2}
\end{equation}
\noindent
where, \textit{m} is the electron mass. $\chi$(\textit{k}) is weighted by \textit{k$^2$} to amplify the oscillation at high \textit{k}  and the \textit{$\chi$(k)k$^2$} functions are subsequently fourier transformed in \textit{R } space to generate the $\chi$(\textit{R}) \textit{vs.} \textit{R} plots in terms of the real distances from the center of the absorbing atom. The ATHENA subroutine available within the Demeter software package~\cite{ravel2005athena} has been used for the above data reduction including background reduction and Fourier transform. Fig.~\ref{EXAFS_FiT_Fig} shows Fourier transformed EXAFS spectra ($\chi$(\textit{R}) \textit{vs.} \textit{R} plots) of the FeMnVGa sample at Fe, Mn, and V edges.

Subsequently, the above experimental $\chi$(\textit{R}) \textit{vs.} \textit{R} data were fitted with theoretically generated plots. The structural parameters (lattice parameters and atomic coordination numbers) acquired from the XRD/ND data have been used for theoretical modelling of the EXAFS spectra of FeMnVGa. During the fitting, bond distances (\textit{R}), co-ordination numbers (\textit{N}) (including scattering amplitudes) and disorder (Debye-Waller) factors (${\sigma^2}$), which give the mean square fluctuations in the distances, have been used as fitting parameters. In this work,  $\chi$(\textit{R}) \textit{vs.} \textit{R} plots measured at Fe, Mn, and V edges are ﬁtted simultaneously with common ﬁtting parameters and for all the edges data ﬁtting was done for a distance up to 5 {\AA}.

The goodness of fit has been determined by the value of the \textit{R}$_{factor}$  defined by:
\begin{equation}
 \scalebox{1} {$R_{factor}= \frac{[Im(\chi_{dat}(r_{i})-\chi_{th}(r_{i})]^{2} + [Re((\chi_{dat}(r_{i})-\chi_{th}(r_{i})]^{2}]^{2}} {[Im(\chi_{dat}(r_i)^{2}]+[Re(\chi_{dat}(r_i)^{2}]}$}
\label{eq3}
\end{equation}
\noindent
where, ${\chi_{dat}}$ and ${\chi_{th}}$ refer to the experimental and theoretical  values respectively and \textit{Im} and \textit{Re} refer to the imaginary and real parts of the respective quantities. The subroutines ATOMS and ARTEMIS available within the Demeter software package~\cite{ravel2005athena} have been used respectively for generation of the theoretical paths from the crystallographic structure and fitting of the experimental data with the theoretical simulation. Fig.~\ref{EXAFS_FiT_Fig} shows the best fit theoretical spectra along with the experimental data and the best fit parameters have been given in Table-\ref{Tab:EXAFS}.

\begin{table*}[ht]
\caption{Bond length (R), coordination number (N), and Debye-Waller or disorder factor (${\sigma}^2$) obtained by EXAFS fitting for FeMnVGa at Fe, Mn and V edge.}

\begin{tabular}{cccccccccccc}
\hline\hline
\multicolumn{4}{c}{Fe edge} & \multicolumn{4}{c}{Mn edge} & \multicolumn{4}{c}{V edge}  \\   \hline\hline
 Path & R  ({\AA})  & N & ${\sigma}^2$  & Path & R ({\AA}) & N & ${\sigma}^2$ &Path & R ({\AA}) & N & ${\sigma}^2$   \\ \hline
Fe-Ga&2.48${\pm}$0.01 &4 & 0.0088${\pm}$0.0006 & Mn-Ga &  2.48${\pm}$0.01 & 4 & 0.0073${\pm}$0.0009  & V-Fe  & 2.48${\pm}$0.01  &  4  & 0.0085${\pm}$0.0010                                          \\ \hline
Fe-V &  2.48${\pm}$0.01 & 4  & 0.0175${\pm}$0.0009& Mn-V &  2.48${\pm}$0.01 & 4 &  0.0175${\pm}$0.0009  & V-Mn & 2.48${\pm}$0.01  &  4  &  0.0250${\pm}$0.0081                                      \\ \hline
Fe-Mn&2.87${\pm}$0.03 &  3 & 0.0049${\pm}$0.0044 & Mn-Fe &  2.87${\pm}$0.01 & 2   &  0.0049${\pm}$0.0044  & V-Ga  & 2.87${\pm}$0.01 & 6 &0.0088${\pm}$0.0010                                         \\ \hline
Fe-Fe  & 2.87${\pm}$0.03 &  3 & 0.0174${\pm}$0.0079 &  Mn-Mn & 2.87${\pm}$0.01  & 4   &   0.0166${\pm}$0.0047  &V-V & 4.13${\pm}$0.03 &  12   &   0.0241${\pm}$0.0042                          \\ \hline
Fe-Fe  & 4.06${\pm}$0.03 &  6 & 0.0072${\pm}$0.0013 &  Mn-Mn & 4.06${\pm}$0.01  & 6   &   0.0024${\pm}$0.0022  &V-Mn & 4.86${\pm}$0.02 &  12   &   0.0250${\pm}$0.0240                          \\ \hline
Fe-Mn  & 4.24${\pm}$0.03 &  6 & 0.0075${\pm}$0.0016 &  Mn-Fe & 4.24${\pm}$0.01  & 6   &   0.0033${\pm}$0.0028  &V-Fe & 4.86${\pm}$0.02 &  12   &   0.0070${\pm}$0.0019                        \\ \hline
Fe-Ga  & 4.80${\pm}$0.03 &  12 & 0.0043${\pm}$0.0007 &  Mn-Ga & 4.80${\pm}$0.01  & 12   &   0.0029${\pm}$0.0011  & &  &    &                          \\ \hline
Fe-V  & 4.80${\pm}$0.03 &  12 & 0.0300${\pm}$0.0097 &  Mn-V & 4.80${\pm}$0.01  & 12   &   0.0300${\pm}$0.0146  & &  &    &                              \\ \hline\hline
\end{tabular}
\label{Tab:EXAFS}
\end{table*}

It can be seen from Fig.~\ref{EXAFS_FiT_Fig} that the Fourier transformed spectra of FeMnVGa at all the three edges (Fe, Mn, V) look identical to each other, with a major peak near 2.25 {\AA}. A similar type of EXAFS spectra was earlier observed for CoFeMnGe~\cite{bainsla2015local}. It is also to be noted that simultaneous ﬁtting of all the edges (Fe, Mn, and V) is only possible assuming a disordered structure. In the Fourier transformed spectra of FeMnVGa (Fig.~\ref{EXAFS_FiT_Fig}), for the Fe edge, the ﬁrst major peak near 2.25 {\AA} has contributions from Fe-Ga (2.48 {\AA}), Fe-V (2.48 {\AA}), Fe-Mn (2.87 {\AA}) and Fe-Fe (2.87 {\AA}) paths. The peak near 3.75 {\AA} has contributions from Fe-Fe (4.06 {\AA}) and Fe-Mn (4.24 {\AA}) paths, while the peak near 4.50 {\AA} has contributions from Fe-Ga (4.80 {\AA}) and Fe-V (4.80 {\AA}) paths. Similarly, for the Mn edge, the major peak near 2.25 {\AA} has contributions from Mn-Ga (2.48 {\AA}), Mn-V (2.48 {\AA}), Mn-Fe (2.87 {\AA}) and Mn-Mn (2.87 {\AA}) paths, while for the peak near 3.75 {\AA} has contribution from Mn-Mn (4.06 {\AA}) and Mn-Fe (4.24 {\AA}) paths. Second major peak near 4.5 {\AA} has contributions from Mn-Ga (4.80 {\AA}) and Mn-V (4.80 {\AA}) paths. For the V edge, the ﬁrst major peak near 2.25 {\AA} has contributions from V-Fe (2.48 {\AA}), V-Mn (2.48 {\AA}), and V-Ga (2.87 {\AA}) paths. The small peak near 3.75 {\AA} has a contribution from V-V (4.13 {\AA}), and the peak near 4.5 {\AA} is due to V-Mn (4.86 {\AA}) and V-Fe (4.86 {\AA}) paths. We have also estimated a ratio of mixing of Fe/Mn at 4\textit{c} and 4\textit{d} site from the EXAFS ﬁtting which is found to be 50:50 from the Fe edge. Our EXAFS analysis results thus further support the presence of disordered structure and is consistent with the neutron diffraction results (Sec.\ref{sec:Neutron}) and DFT calculations on the disordered structure discussed later (Sec.\ref{sec:DOS_Disorder}).

\subsection{\label{sec:Magnetism}Magnetic properties}
\begin{figure}[h]
\centerline{\includegraphics[width=.48\textwidth]{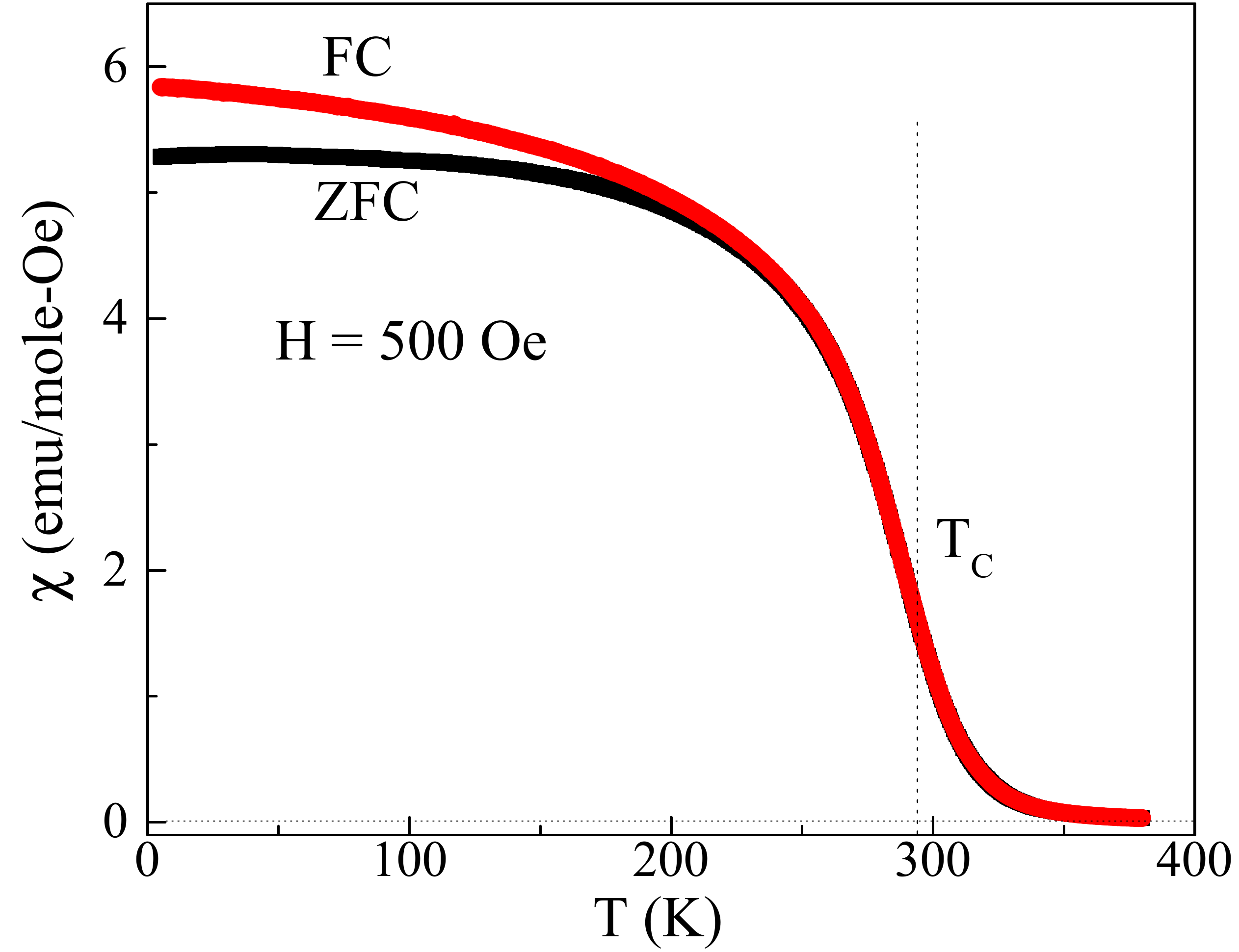}}
{\caption{Temperature dependence of  magnetic susceptibility of FeMnVGa measured in a 500 Oe applied magnetic field under zero field-cooled (ZFC) and field-cooled (FC) configuration. Curie temperature, T$_{\rm C}$, is determined from the minima in d${\chi}$/dT \textit{vs.} T plot.}\label{MT_Fig}}
\end{figure}

\begin{figure}[h]
\centerline{\includegraphics[width=.48\textwidth]{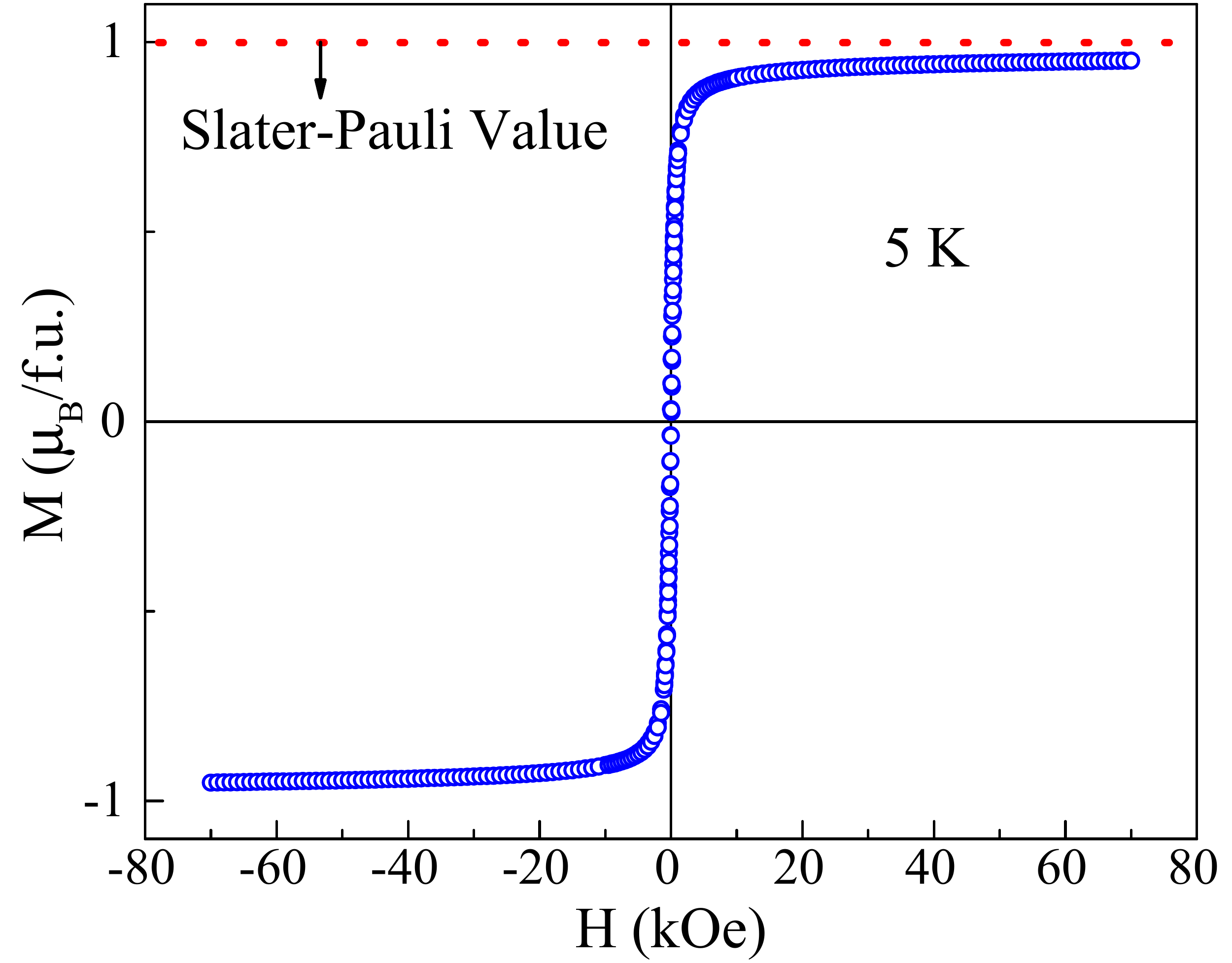}}
{\caption{Isothermal magnetization of FeMnVGa measured at 5 K\@.}\label{MH_Fig}}
\end{figure}

Fig.~\ref{MT_Fig} represents the temperature variation of magnetic susceptibility of FeMnVGa measured at 500 Oe under zero-field cooled (ZFC) and field-cooled (FC) condition. Magnetic susceptibility data clearly show ferromagnetic type behaviour and the transition temperature is determined to be  293 K, obtained from the temperature derivative of the FC magnetic susceptibility data (figure not shown here). The Slater-Pauling (S-P) rule asserts that the total magnetic moment for Heusler alloys is determined by the relation m=(N$_V$ $-$ 24) $\mu_{\rm B}$/f.u., where N$_V$ is the total valence electrons count (VEC) for a material~\cite{galanakis2002slater}. The S-P rule is commonly considered as a useful information for determining the relationship between a compound's magnetism (total spin-magnetic moment) and its electronic structure (the appearance of half-metallicity)~\cite{galanakis2006electronic}. Generally, all the reported Heusler-based HMF are known to follow the S-P rule~\cite{graf2011simple,bainsla2016equiatomic}. For FeMnVGa, the VEC is 23, the absolute value of the S-P moment is 1 $\mu_{\rm B}$/f.u.. The isothermal magnetization of FeMnVGa measured at 5 K (Fig.~\ref{MH_Fig} ) exhibits soft ferromagnetic properties with low hysteresis and the obtained saturation magnetization is to be $\sim$ 0.92 $\mu_{\rm B}$/f.u., which is fairly close to the value predicted by S-P rule and DFT calculations.

\subsection{\label{sec:Mossbauer}M\"{o}ssbauer spectrometry}
\begin{figure}[h]
\centerline{\includegraphics[width=.48\textwidth]{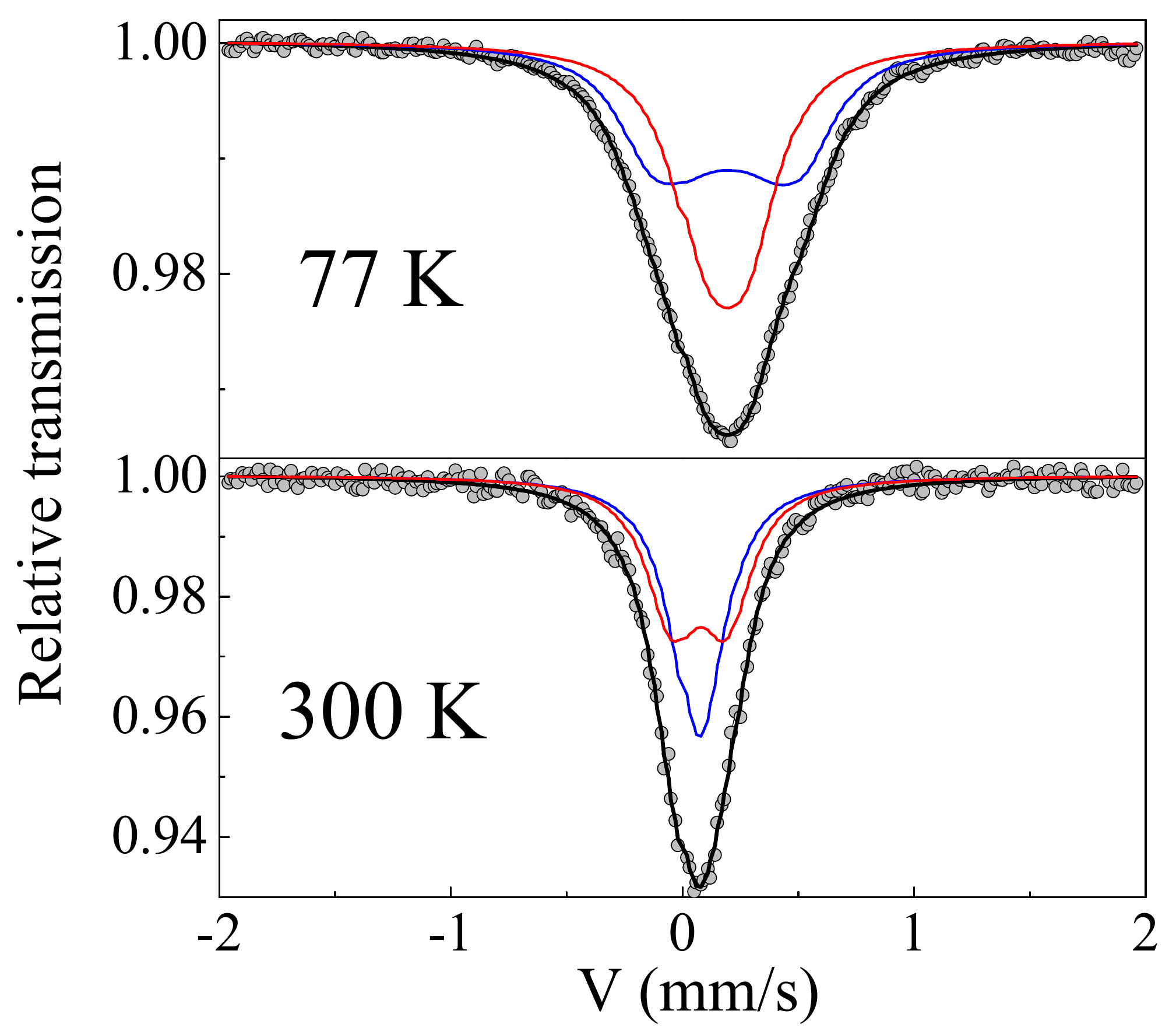}}
{\caption{M\"{o}ssbauer spectra of FeMnVGa taken at (bottom) 300 K and (bottom) 77 K (top).}\label{Fig_Mossbauer}}
\end{figure}

$^{57}$Fe M\"{o}ssbauer measurements at 300 K and 77 K were carried out to obtain further insight of the structural disorder and magnetism in FeMnVGa, in relation to the Fe atom. In Fig.~\ref{Fig_Mossbauer}, for both the temperatures, there is just one widened and slightly asymmetrical line in the spectrum, and it is not of Lorentzian shape. Based on the Fe atom's nearly-cubic symmetric surroundings, the 300 K spectrum may be explained by at least two quadrupolar doublets with tiny quadrupolar splitting values as shown in Fig.~\ref{Fig_Mossbauer}  and Table~\ref{Mossbauer_Table}. At 77 K, the spectrum is widened further and  may be represented by two quadrupolar components. A slight increase in quadrupolar strength may be perceived as due to the development of a hyperfine field which is weak enough to be properly convoluted due to poor resolution. For the simplicity of analysis, the spectrum can be characterised by two components of almost equal intensity. Lower values of hyperfine fields (7 kOe and 20 kOe at the two Fe sites , respectively) suggest that Fe atoms do not participate in ferromagnetic ordering, indicating that the Mn moments are the primary contributor to most of the total magnetic moment. This is perfectly consistent with neutron diffraction (Sec.\ref{sec:Neutron}) and DFT-based theoretical models (Sec.\ref{sec:DOS_Disorder}) discussed later. One may note here that although Fe has just one crystallographic site in Type-2 ordered structure, the existence in the M\"{o}ssbauer spectra of two components of almost equal intensities and similar hyperfine parameters suggests the availability of two locations for Fe atoms with nearly analogous structural environments. The crystal symmetry of quaternary Heusler alloy (space group: \textit{F$\bar{4}$3m}) is such that all the 4\textit{a}, 4\textit{b}, 4\textit{c} and 4\textit{d} sites correspond to tetrahedral symmetry and cannot be distinguished, leaving four options for the two Fe-sites. Nonetheless, in the disordered Type-2 structure where Fe and Mn randomly share the same sites (Sec.\ref{sec:Neutron}), the 4\textit{c} and 4\textit{d} sites become equivalent and the structure can then be described by space group \textit{Fm$\bar{3}$m}. The 4\textit{c} and 4\textit{d} sites of \textit{F$\bar{4}$3m} merge into the 8\textit{c} site of Fm$\bar{3}$m with tetrahedral symmetry, whereas the 4\textit{a }and 4\textit{b} sites of the latter space group display a distinguishable octahedral symmetry. By suggesting that the two locations for Fe atoms display analogous structural environments, M\"{o}ssbauer thus confirms the structural model derived from neutron diffraction. The two components are a priori rather equal, but the whole hyperfine structure is not well resolved, preventing a physically accurate estimation.

\begin{table}[]
\caption{ Fitted parameter values for the M\"{o}ssbauer spectra of FeMnVGa. Isomer shift ($\delta$) , line-width at half height (${\Gamma}$) (quoted relative to $\alpha$-Fe at 300 K), quadrupolar shift (${\frac {Q} {2\varepsilon}}$), hyperfine field (B$_{hf}$) and relative proportions ($\%$) are estimated at 300 K and 77 K.}
\begin{tabular}{ccccccc}
\hline
\hline
 T (K) & Site & $\delta$ (mm/s) & ${\Gamma}$ (mm/s) & ${\frac {Q} {2\varepsilon}}$  & B$_{hf} (T) $  & $\%$     \\
 & & $\pm$0.01 &$\pm$0.01 &$\pm$0.01 &$\pm$0.3 &$\pm$2\\ \hline
300    & Fe1 & 0.19 & 0.26 & 0.00  & - & 50                         \\
           & Fe2 & 0.19 & 0.26 & 0.22  & - & 50                         \\
77 K  & Fe1 & 0.31 & 0.34 & 0.00  &0.7 & 50                         \\
         & Fe2 & 0.31 & 0.32 & -0.01  & 2.0& 50                         \\

\hline
\end{tabular}
\label{Mossbauer_Table}
\end{table}

\subsection{\label{sec:Magnetic_Structure}Magnetic Structure}

\begin{figure}[h]
\centerline{\includegraphics[width=.48\textwidth]{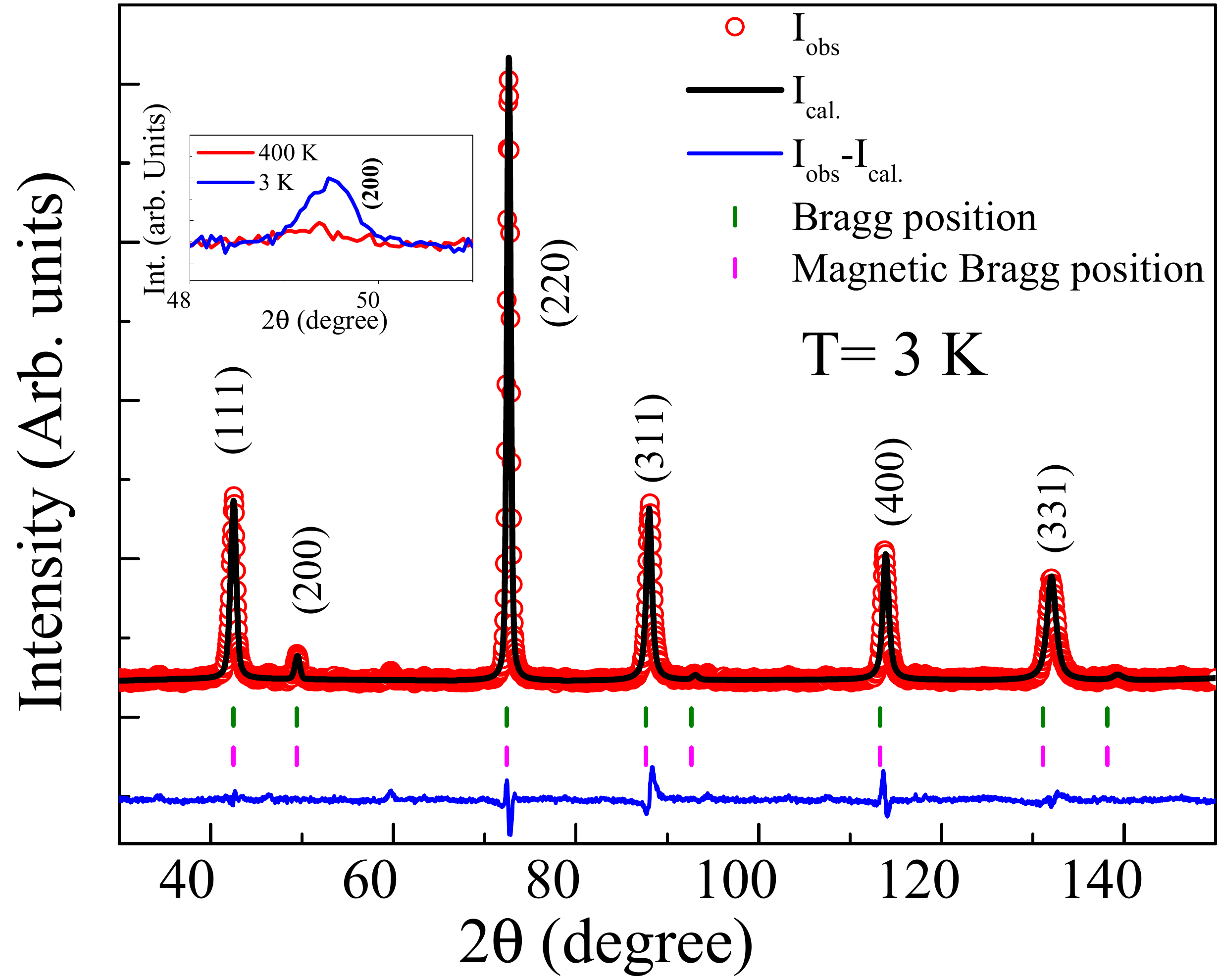}}
{\caption{Rietveld refinement of the Neutron diffraction pattern of FeMnVGa taken at 3 K. The inset in (b) shows the enhancement of (200) Bragg peak intensity due to ferromagnetic ordering.}\label{NPD3K_Fig}}
\end{figure}
Neutron diffraction measurement in the magnetically ordered state can also help us in understanding the magnetic spin arrangement in the system. For this purpose, we have performed the neutron diffraction experiment at 3 K and represented in Fig~\ref{NPD3K_Fig}, much below the magnetic transition temperature of 293 K. For antiferromagnetic (AFM) materials, below their ordering temperature, additional Bragg peaks are observed in the neutron diffraction pattern. In contrast to AFM, no additional peaks are seen in the diffraction pattern of a ferromagnetic lattice, although some of the existing Bragg peaks become more intense below T$_{\rm C}$. For FeMnVGa, a clear increase of (200) Bragg peak is overtly evident (inset of Fig.~\ref{NPD3K_Fig}) which confirms the occurrence of long-range FM ordering, assuming the structural model obtained from the paramagnetic region (400 K) remain invariant. Rietveld refinement of the 3 K data was performed with an additional consideration of a magnetic structure. Given the cubic symmetry of the crystal structure of FeMnVGa, powder neutron diffraction cannot determine the direction of the moments in a FM structure. The obtained total magnetic moment is found to be 0.81(2) $\mu_{\rm B}$/f.u. for Mn atoms, which closely matches  with the moment obtained from the isothermal magnetization (0.92 $\mu_{\rm B}$/f.u. measured at 5 K). The neutron diffraction measurement thus confirms that Mn is the major contributor to observed magnetism in the studied compound, consistent with the M\"{o}ssbauer analysis which unveiled an absence (low value) of hyperfine field for Fe atoms (Sec.\ref{sec:Mossbauer}) and theoretical calculations (Sec.\ref{sec:DOS_Disorder}). DFT calculations also estimated a site specific moment of 0.53 $\mu_{\rm B}$/f.u. for the V atoms. However, neutron diffraction discard any possibility of long-range order of V atom. The moment on the V-site can be best considered as a induced moment, induced by the magnetically ordered Mn-spin. Theoretical estimates of the magnetic moment often considered as qualitative descriptions because they do not always quantitatively correlate to the empirically obtained value.
\subsection{\label{sec:DOS_Disorder}Electronic structure calculations--Disordered structure}

\begin{figure}[h]
\centerline{\includegraphics[width=.48\textwidth]{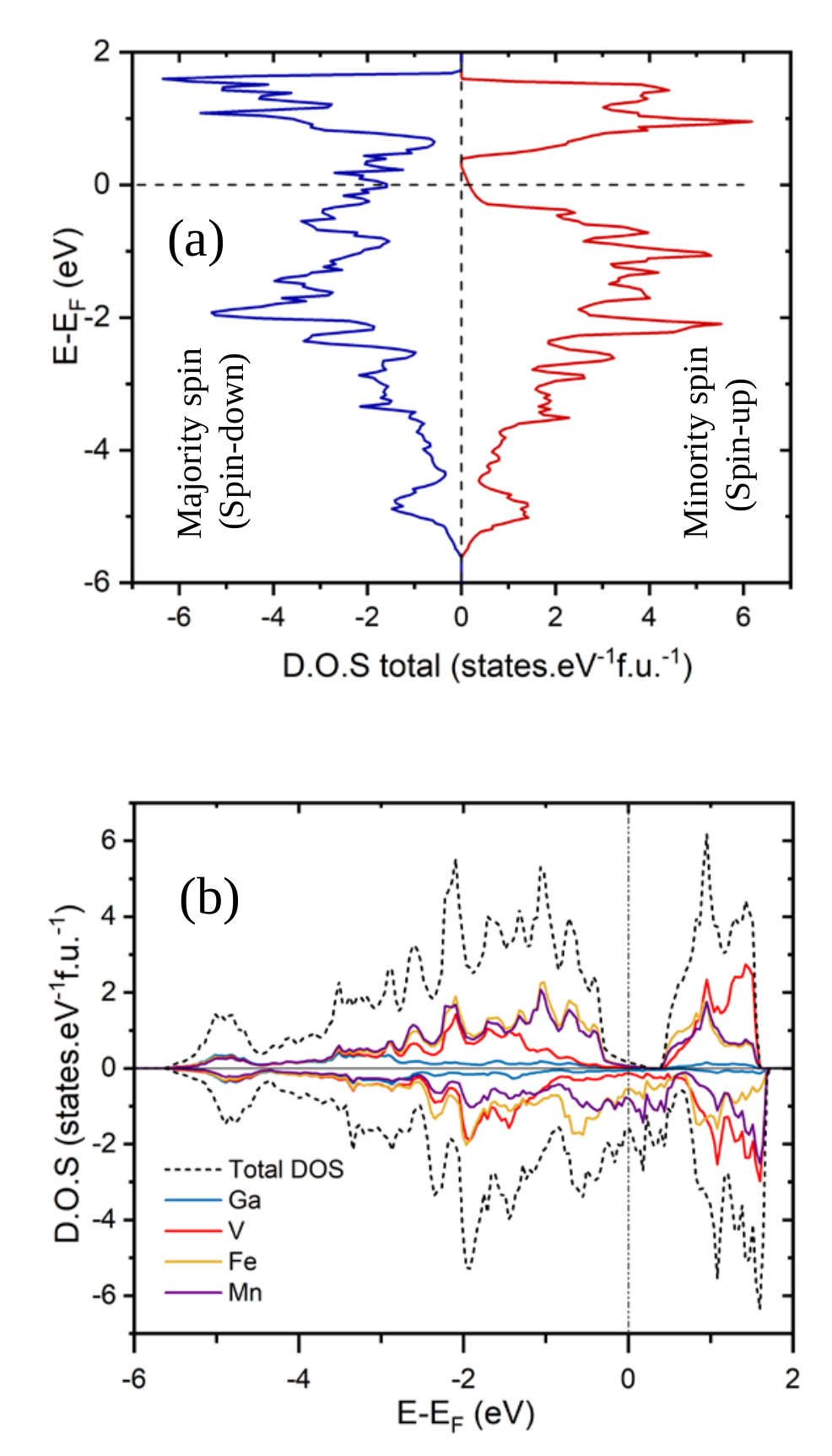}}
{\caption{(a) Density of states of  FeMnVGa in disordered structure (b) Electronic DOS (total and partial) of disordered FeMnVGa.}\label{DOS_Disorder}}
\end{figure}

When considering the crystal structure of FeMnVGa, the most plausible scenario is a disordered structure with Fe and Mn evenly distributed among 4$c$ and 4$d$ sites, as determined by Neutron diffraction. As a result, we have revised the electronic structure analysis to account for the system's disorder. Subsequently, we find that the SQS disordered structure's enthalpy of formation ($\Delta_f{H}$) is indeed lower than that of the ordered Type-2 structure, with a value of -28.58\,{kJ/mol}, \textit{i.e.} the formation energy is 1.37\,{kJ/mol} lower for the disordered structure \textit{vis-\`{a}-vis} the ordered structure. Fig.~\ref{DOS_Disorder} represents the spin polarized density of states (DOS) of the disordered structure. The band gap is maintained by the Fermi level at the minority (spin-up) band, meaning that disordered FeMnVGa preserves its half-metallic ferromagnetic ground state.  Despite the strong atomic disorder, the robustness of half-metallicity is evident as the spin-polarisation value is changed only marginally (P = 81.3\%) in the disordered structure. This is due to strange fact that disorder (between 4\textit{c} and 4\textit{d} sites) enhances the symmetry of the crystal structure from space group: \textit{F$\bar{4}$3m} (no. 216) to space group : \textit{Fm$\bar{3}$m} (no. 225) conserving the main features (half-metallic ferromagnetic) of the electronic structure. The total magnetic contribution in the disordered structure (-0.89 $\mu_B$/f.u.) basically remains constant in comparison to the ordered structure, as it is in the case of polarisation. Even though the magnetic contributions of V and Ga to the total magnetic moment remain equal to the ordered structure, V = 0.50 $\mu_B$/f.u. and Ga = 0.02 $\mu_B$/f.u., the contribution of Mn to the magnetic moment grows substantially at the expense of Fe: Fe = -0.21 $\mu_B$/f.u., Mn = -1.2 $\mu_B$/f.u. Mn spins at the 4\textit{c} and 4\textit{d} positions coupled ferromagnetically in the disordered structure. It should be mentioned that we have also examined the possibility of anti-ferromagnetic coupling of the Mn spins. However, the corresponding energy converges to the same value as that of ferromagnetic coupling and further validates the ferromagnetic spin arrangements of the Mn atoms.

\subsection{\label{sec:Resistivity}Resistivity}

\begin{figure}[h]
\centerline{\includegraphics[width=.48\textwidth]{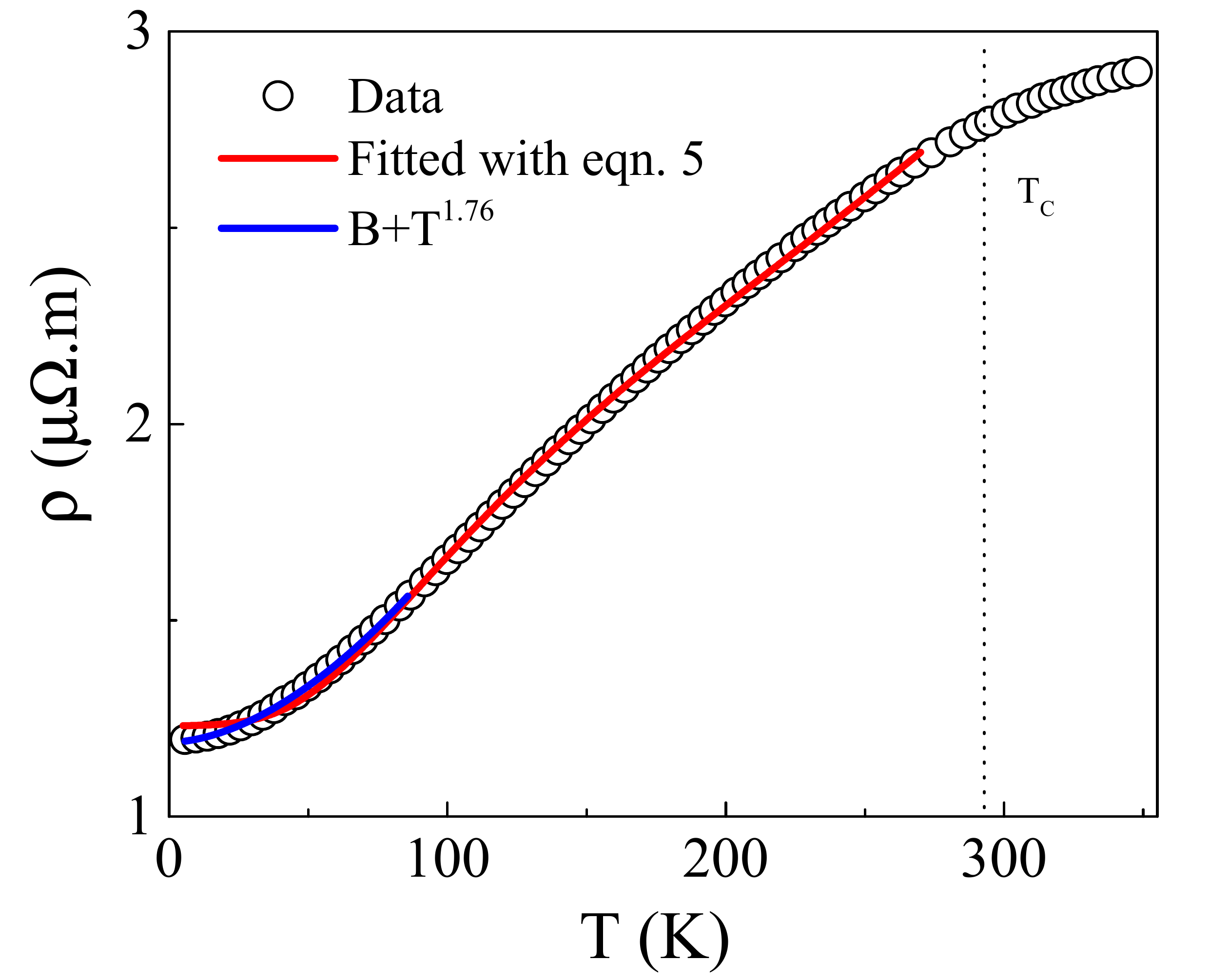}}
{\caption{Temperature dependence of the electrical resistivity measured in the absence of magnetic field in the temperature range 5$-$350 K.}\label{RT_Fig}}
\end{figure}

To find the signature of HMF state in FeMnVGa, temperature variation of the resistivity ($\rho$(T)) was measured in the regime 5$-$350 K (Fig.~\ref{RT_Fig}). Throughout the measurement range, the $\rho$(T) data displays metal-like behaviour. Near the magnetic transition temperature, however, a blunt change in slope was detected. The residual resistivity ratio (RRR={{$\rho_{350 {\rm K}}$}}/{{$\rho_{5 {\rm K}}$}}) was estimated to be 2.45. The low value of the RRR points toward the presence of structural disorder which earlier has been confirmed by M\"{o}ssbauer, Neutron and EXAFS measurements~\cite{rani2017structural,boeuf2006low,nag2022cofevsb,samanta2020structural}. We attempted to fit the resistivity data in the magnetically ordered state (5 $<$ T $<$ 270 K) using the following formula~\cite{gui2021ferromagnetic,rossiter1991electrical},

\begin{equation}
\rho(T) = \rho_0 + A{\bigg(\frac{T}{\Theta_{D}}\bigg)}^5 \int_{0}^{\frac{\Theta_{D}}{T}} \frac{x^{5}}{(e^{x}-1)(1-e^{-x})} dx +BT^2
\label{eqRes1}
\end{equation}
\noindent
where $\rho_0$ is the residual resistivity that arises due to lattice defects, irregularities, \textit{etc}. and the second term provides the phonon scattering mechanism, in which, \textit{A} and $\Theta_{D}$ are phonon scattering constant and Debye constant~\cite{gruneisen1933abhangigkeit}, respectively. The last term, \textit{B}T$^2$, is due to magnon scattering, which remains only up to T$_{\rm C}$~\cite{bombor2013half}. It may be noted here that in most of the known HMF systems, the magnon contribution is reported to be quite small, in comparison to phonon contribution~\cite{bombor2013half,rani2017structural,bainsla2016equiatomic}. The same is true in the present system, FeMnVGa, as well.

Eqn.~\ref{eqRes1}, however, could not fit the experimental data well in the low temperature region 5$-$85 K. Instead, low temperature (5$-$85 K) data could be well fitted using the following equation,
\begin{equation}
\rho= B+CT^n
\label{eqRes2}
\end{equation}
 and the value of \textit{n} is found to be 1.76, which can not be associated with any known kind of scattering. Eqn.~\ref{eqRes2} is widely used to fit the resistivity data for Heusler-based HMF systems in the low temperature region. Different values of the exponent \textit{n} are reported in literature in many HMF systems~\cite{rani2017structural,samanta2020structural,bainsla2015high,rani2019experimental}. For a simple ferromagnetic (FM) material, a quadratic temperature dependence of $\rho$(T) is expected due to the magnon scattering. This value of \textit{n} not equal to 2 signifies the absence of magnon scattering in the magnetically ordered region, which may look surprising at first sight. However, it is worth to mention here that in contrast to the standard FM compounds, the HMF systems possess only one kind of band (majority or minority) at the Fermi level (E$_{\rm F}$). Magnon contribution, which primarily associated with spin flip scattering , therefore may not able to contribute in a HMF system due to absence of one kind of band at E$_{\rm F}$. Thus, our resistivity data also provide an  indirect support to the presence of HMF state in FeMnVGa.

\section{Conclusion}
In summary, we have reported here the formation of a Fe-based novel quaternary polycrystalline Heusler alloy FeMnVGa in single phase using arc-melting technique. Neutron and X-ray diffraction, EXAFS, ${^{57}}$Fe M\"{o}ssbauer spectrometry results, together with DFT calculation confirms that the compound crystalizes in Type-2 disordered structure in which Ga is at 4\textit{a}, V at 4\textit{b} and Fe/Mn at 4\textit{c}/4\textit{d} with 50:50 proportions. Magnetic susceptibility unambiguously  exhibits ferromagnetic transistion near 293 K. The saturation magnetic moment obtained from isothermal magnetization is found to be 0.92 $\mu_{\rm B}$/f.u. which is very close to the value 1 $\mu_{\rm B}$/f.u. predicted by the Slater-Pauling rule. M\"{o}ssbauer spectrometry, along with neutron diffraction suggest that Mn is the major contributor to the total magnetism in the compound consistent with the theoretical calculations. DFT calculations on the disordered structure reveal that the HMF state persists even in the disordered structure, as the spin polarization merely changes from 89.9 \% to 81.3 \% as a result of this disorder. Such robustness of HMF behaviour despite having a large crystal disorder are rarely displayed by the Heusler alloy group of material and is quite unique over other reported HMF materials in Heusler alloys. Incidentally, such coexistence of high spin-polarization along with structural disorder have earlier been reported in a related system, FeMnVAl~\cite{gupta2022coexisting}. In both the cases, structural disorder enhances the symmetry of the crystal structure maintaining it’s prime feature (half-metallcity) near the Fermi level. Robust half-metallicity has also been theoretically estimated for different types of site-disorders in CoFeMnSi~\cite{feng2015effect}, some of which even lower the crystal symmetry. Although the spin-polarization remain high in most of these predicted configurations, the large band gap is however maintained only in the configuration where the anti-site disorder involves the swapping of Co and Fe in 50:50 ratio that improves crystal structural symmetry from \textit{F$\bar{4}$3m} to \textit{Fm$\bar{3}$m}, which is also consistent with our observations. One may thus tempt to correlated the high spin polarisation with disorder-induced symmetry enhancement. On the other hand, systems like MnCrVAl forms with A2-type disorder~\cite{kharel2017effect}, which happens to have higher symmetry than the ordered \textit{Y}-type structure,  the material lacks the theoretically anticipated SGS characteristics. Thus, it may not be possible to draw a universal rule correlating the high spin-polarization with disorder-induced symmetry enhancement for any arbitrary type of disorder. Due to the introduction of structural disorder in the thin films or devices based on HMFs materials, existing reported HMF exhibit very low polarization when compared to their poly-/single-crystalline mother compound. In such cases, a compound like FeMnVGa, whose HMF property is very robust or unaffected by structural disorder, would be very useful in the preparation of thin films and devices for spintronic applications.

\section{Acknowledgement}
S.G and S.C would like to sincerely acknowledge SINP, India and UGC, India, respectively, for their fellowship. DFT calculations were performed using HPC resources from GENCI-CINES (Grant 2021-A0100906175). Work at the Ames Laboratory (in part) was supported by the Department of Energy- Basic Energy Sciences, Materials Sciences and Engineering Division, under Contract No. DE-AC02-07CH11358. 
\normalem
\bibliographystyle{apsrev4-2}

\begin{thebibliography}{57}%
\makeatletter
\providecommand \@ifxundefined [1]{%
 \@ifx{#1\undefined}
}%
\providecommand \@ifnum [1]{%
 \ifnum #1\expandafter \@firstoftwo
 \else \expandafter \@secondoftwo
 \fi
}%
\providecommand \@ifx [1]{%
 \ifx #1\expandafter \@firstoftwo
 \else \expandafter \@secondoftwo
 \fi
}%
\providecommand \natexlab [1]{#1}%
\providecommand \enquote  [1]{``#1''}%
\providecommand \bibnamefont  [1]{#1}%
\providecommand \bibfnamefont [1]{#1}%
\providecommand \citenamefont [1]{#1}%
\providecommand \href@noop [0]{\@secondoftwo}%
\providecommand \href [0]{\begingroup \@sanitize@url \@href}%
\providecommand \@href[1]{\@@startlink{#1}\@@href}%
\providecommand \@@href[1]{\endgroup#1\@@endlink}%
\providecommand \@sanitize@url [0]{\catcode `\\12\catcode `\$12\catcode
  `\&12\catcode `\#12\catcode `\^12\catcode `\_12\catcode `\%12\relax}%
\providecommand \@@startlink[1]{}%
\providecommand \@@endlink[0]{}%
\providecommand \url  [0]{\begingroup\@sanitize@url \@url }%
\providecommand \@url [1]{\endgroup\@href {#1}{\urlprefix }}%
\providecommand \urlprefix  [0]{URL }%
\providecommand \Eprint [0]{\href }%
\providecommand \doibase [0]{https://doi.org/}%
\providecommand \selectlanguage [0]{\@gobble}%
\providecommand \bibinfo  [0]{\@secondoftwo}%
\providecommand \bibfield  [0]{\@secondoftwo}%
\providecommand \translation [1]{[#1]}%
\providecommand \BibitemOpen [0]{}%
\providecommand \bibitemStop [0]{}%
\providecommand \bibitemNoStop [0]{.\EOS\space}%
\providecommand \EOS [0]{\spacefactor3000\relax}%
\providecommand \BibitemShut  [1]{\csname bibitem#1\endcsname}%
\let\auto@bib@innerbib\@empty
\bibitem [{\citenamefont {Felser}\ \emph {et~al.}(2007)\citenamefont {Felser},
  \citenamefont {Fecher},\ and\ \citenamefont {Balke}}]{felser2007spintronics}%
  \BibitemOpen
  \bibfield  {author} {\bibinfo {author} {\bibfnamefont {C.}~\bibnamefont
  {Felser}}, \bibinfo {author} {\bibfnamefont {G.~H.}\ \bibnamefont {Fecher}},\
  and\ \bibinfo {author} {\bibfnamefont {B.}~\bibnamefont {Balke}},\
  }\href@noop {} {\bibfield  {journal} {\bibinfo  {journal} {Angew. Chem. Int.
  Ed.}\ }\textbf {\bibinfo {volume} {46}},\ \bibinfo {pages} {668} (\bibinfo
  {year} {2007})}\BibitemShut {NoStop}%
\bibitem [{\citenamefont {Wolf}\ \emph {et~al.}(2001)\citenamefont {Wolf},
  \citenamefont {Awschalom}, \citenamefont {Buhrman}, \citenamefont {Daughton},
  \citenamefont {von Moln{\'a}r}, \citenamefont {Roukes}, \citenamefont
  {Chtchelkanova},\ and\ \citenamefont {Treger}}]{wolf2001spintronics}%
  \BibitemOpen
  \bibfield  {author} {\bibinfo {author} {\bibfnamefont {S.~A.}\ \bibnamefont
  {Wolf}}, \bibinfo {author} {\bibfnamefont {D.~D.}\ \bibnamefont {Awschalom}},
  \bibinfo {author} {\bibfnamefont {R.}~\bibnamefont {Buhrman}}, \bibinfo
  {author} {\bibfnamefont {J.}~\bibnamefont {Daughton}}, \bibinfo {author}
  {\bibfnamefont {v.~S.}\ \bibnamefont {von Moln{\'a}r}}, \bibinfo {author}
  {\bibfnamefont {M.}~\bibnamefont {Roukes}}, \bibinfo {author} {\bibfnamefont
  {A.~Y.}\ \bibnamefont {Chtchelkanova}},\ and\ \bibinfo {author}
  {\bibfnamefont {D.}~\bibnamefont {Treger}},\ }\href@noop {} {\bibfield
  {journal} {\bibinfo  {journal} {Science}\ }\textbf {\bibinfo {volume}
  {294}},\ \bibinfo {pages} {1488} (\bibinfo {year} {2001})}\BibitemShut
  {NoStop}%
\bibitem [{\citenamefont {De~Groot}\ \emph {et~al.}(1983)\citenamefont
  {De~Groot}, \citenamefont {Mueller}, \citenamefont {Van~Engen},\ and\
  \citenamefont {Buschow}}]{de1983new}%
  \BibitemOpen
  \bibfield  {author} {\bibinfo {author} {\bibfnamefont {R.~A.}\ \bibnamefont
  {De~Groot}}, \bibinfo {author} {\bibfnamefont {F.~M.}\ \bibnamefont
  {Mueller}}, \bibinfo {author} {\bibfnamefont {P.~G.}\ \bibnamefont
  {Van~Engen}},\ and\ \bibinfo {author} {\bibfnamefont {K.}~\bibnamefont
  {Buschow}},\ }\href@noop {} {\bibfield  {journal} {\bibinfo  {journal} {Phys.
  Rev. Lett.}\ }\textbf {\bibinfo {volume} {50}},\ \bibinfo {pages} {2024}
  (\bibinfo {year} {1983})}\BibitemShut {NoStop}%
\bibitem [{\citenamefont {Park}\ \emph {et~al.}(1998)\citenamefont {Park},
  \citenamefont {Vescovo}, \citenamefont {Kim}, \citenamefont {Kwon},
  \citenamefont {Ramesh},\ and\ \citenamefont {Venkatesan}}]{park1998direct}%
  \BibitemOpen
  \bibfield  {author} {\bibinfo {author} {\bibfnamefont {J.-H.}\ \bibnamefont
  {Park}}, \bibinfo {author} {\bibfnamefont {E.}~\bibnamefont {Vescovo}},
  \bibinfo {author} {\bibfnamefont {H.-J.}\ \bibnamefont {Kim}}, \bibinfo
  {author} {\bibfnamefont {C.}~\bibnamefont {Kwon}}, \bibinfo {author}
  {\bibfnamefont {R.}~\bibnamefont {Ramesh}},\ and\ \bibinfo {author}
  {\bibfnamefont {T.}~\bibnamefont {Venkatesan}},\ }\href@noop {} {\bibfield
  {journal} {\bibinfo  {journal} {Nature}\ }\textbf {\bibinfo {volume} {392}},\
  \bibinfo {pages} {794} (\bibinfo {year} {1998})}\BibitemShut {NoStop}%
\bibitem [{\citenamefont {Jourdan}\ \emph {et~al.}(2014)\citenamefont
  {Jourdan}, \citenamefont {Min{\'a}r}, \citenamefont {Braun}, \citenamefont
  {Kronenberg}, \citenamefont {Chadov}, \citenamefont {Balke}, \citenamefont
  {Gloskovskii}, \citenamefont {Kolbe}, \citenamefont {Elmers}, \citenamefont
  {Sch{\"o}nhense} \emph {et~al.}}]{jourdan2014direct}%
  \BibitemOpen
  \bibfield  {author} {\bibinfo {author} {\bibfnamefont {M.}~\bibnamefont
  {Jourdan}}, \bibinfo {author} {\bibfnamefont {J.}~\bibnamefont {Min{\'a}r}},
  \bibinfo {author} {\bibfnamefont {J.}~\bibnamefont {Braun}}, \bibinfo
  {author} {\bibfnamefont {A.}~\bibnamefont {Kronenberg}}, \bibinfo {author}
  {\bibfnamefont {S.}~\bibnamefont {Chadov}}, \bibinfo {author} {\bibfnamefont
  {B.}~\bibnamefont {Balke}}, \bibinfo {author} {\bibfnamefont
  {A.}~\bibnamefont {Gloskovskii}}, \bibinfo {author} {\bibfnamefont
  {M.}~\bibnamefont {Kolbe}}, \bibinfo {author} {\bibfnamefont {H.-J.}\
  \bibnamefont {Elmers}}, \bibinfo {author} {\bibfnamefont {G.}~\bibnamefont
  {Sch{\"o}nhense}}, \emph {et~al.},\ }\href@noop {} {\bibfield  {journal}
  {\bibinfo  {journal} {Nat. Commun}\ }\textbf {\bibinfo {volume} {5}},\
  \bibinfo {pages} {1} (\bibinfo {year} {2014})}\BibitemShut {NoStop}%
\bibitem [{\citenamefont {Bombor}\ \emph {et~al.}(2013)\citenamefont {Bombor},
  \citenamefont {Blum}, \citenamefont {Volkonskiy}, \citenamefont {Rodan},
  \citenamefont {Wurmehl}, \citenamefont {Hess},\ and\ \citenamefont
  {B{\"u}chner}}]{bombor2013half}%
  \BibitemOpen
  \bibfield  {author} {\bibinfo {author} {\bibfnamefont {D.}~\bibnamefont
  {Bombor}}, \bibinfo {author} {\bibfnamefont {C.~G.~F.}\ \bibnamefont {Blum}},
  \bibinfo {author} {\bibfnamefont {O.}~\bibnamefont {Volkonskiy}}, \bibinfo
  {author} {\bibfnamefont {S.}~\bibnamefont {Rodan}}, \bibinfo {author}
  {\bibfnamefont {S.}~\bibnamefont {Wurmehl}}, \bibinfo {author} {\bibfnamefont
  {C.}~\bibnamefont {Hess}},\ and\ \bibinfo {author} {\bibfnamefont
  {B.}~\bibnamefont {B{\"u}chner}},\ }\href@noop {} {\bibfield  {journal}
  {\bibinfo  {journal} {Phys. Rev. Lett.}\ }\textbf {\bibinfo {volume} {110}},\
  \bibinfo {pages} {066601} (\bibinfo {year} {2013})}\BibitemShut {NoStop}%
\bibitem [{\citenamefont {Graf}\ \emph {et~al.}(2011)\citenamefont {Graf},
  \citenamefont {Felser},\ and\ \citenamefont {Parkin}}]{graf2011simple}%
  \BibitemOpen
  \bibfield  {author} {\bibinfo {author} {\bibfnamefont {T.}~\bibnamefont
  {Graf}}, \bibinfo {author} {\bibfnamefont {C.}~\bibnamefont {Felser}},\ and\
  \bibinfo {author} {\bibfnamefont {S.~S.~P.}\ \bibnamefont {Parkin}},\
  }\href@noop {} {\bibfield  {journal} {\bibinfo  {journal} {Prog. Solid. State
  Ch.}\ }\textbf {\bibinfo {volume} {39}},\ \bibinfo {pages} {1} (\bibinfo
  {year} {2011})}\BibitemShut {NoStop}%
\bibitem [{\citenamefont {Ouardi}\ \emph {et~al.}(2013)\citenamefont {Ouardi},
  \citenamefont {Fecher}, \citenamefont {Felser},\ and\ \citenamefont
  {K{\"u}bler}}]{ouardi2013realization}%
  \BibitemOpen
  \bibfield  {author} {\bibinfo {author} {\bibfnamefont {S.}~\bibnamefont
  {Ouardi}}, \bibinfo {author} {\bibfnamefont {G.~H.}\ \bibnamefont {Fecher}},
  \bibinfo {author} {\bibfnamefont {C.}~\bibnamefont {Felser}},\ and\ \bibinfo
  {author} {\bibfnamefont {J.}~\bibnamefont {K{\"u}bler}},\ }\href@noop {}
  {\bibfield  {journal} {\bibinfo  {journal} {Phys. Rev. Lett.}\ }\textbf
  {\bibinfo {volume} {110}},\ \bibinfo {pages} {100401} (\bibinfo {year}
  {2013})}\BibitemShut {NoStop}%
\bibitem [{\citenamefont {Bainsla}\ \emph
  {et~al.}(2015{\natexlab{a}})\citenamefont {Bainsla}, \citenamefont {Mallick},
  \citenamefont {Raja}, \citenamefont {Nigam}, \citenamefont {Varaprasad},
  \citenamefont {Takahashi}, \citenamefont {Alam}, \citenamefont {Suresh},\
  and\ \citenamefont {Hono}}]{bainsla2015spin}%
  \BibitemOpen
  \bibfield  {author} {\bibinfo {author} {\bibfnamefont {L.}~\bibnamefont
  {Bainsla}}, \bibinfo {author} {\bibfnamefont {A.~I.}\ \bibnamefont
  {Mallick}}, \bibinfo {author} {\bibfnamefont {M.~M.}\ \bibnamefont {Raja}},
  \bibinfo {author} {\bibfnamefont {A.~K.}\ \bibnamefont {Nigam}}, \bibinfo
  {author} {\bibfnamefont {B.~S. D. C.~S.}\ \bibnamefont {Varaprasad}},
  \bibinfo {author} {\bibfnamefont {Y.~K.}\ \bibnamefont {Takahashi}}, \bibinfo
  {author} {\bibfnamefont {A.}~\bibnamefont {Alam}}, \bibinfo {author}
  {\bibfnamefont {K.~G.}\ \bibnamefont {Suresh}},\ and\ \bibinfo {author}
  {\bibfnamefont {K.}~\bibnamefont {Hono}},\ }\href@noop {} {\bibfield
  {journal} {\bibinfo  {journal} {Phys. Rev. B}\ }\textbf {\bibinfo {volume}
  {91}},\ \bibinfo {pages} {104408} (\bibinfo {year}
  {2015}{\natexlab{a}})}\BibitemShut {NoStop}%
\bibitem [{\citenamefont {Bainsla}\ \emph
  {et~al.}(2015{\natexlab{b}})\citenamefont {Bainsla}, \citenamefont {Mallick},
  \citenamefont {Raja}, \citenamefont {Coelho}, \citenamefont {Nigam},
  \citenamefont {Johnson}, \citenamefont {Alam},\ and\ \citenamefont
  {Suresh}}]{bainsla2015origin}%
  \BibitemOpen
  \bibfield  {author} {\bibinfo {author} {\bibfnamefont {L.}~\bibnamefont
  {Bainsla}}, \bibinfo {author} {\bibfnamefont {A.~I.}\ \bibnamefont
  {Mallick}}, \bibinfo {author} {\bibfnamefont {M.~M.}\ \bibnamefont {Raja}},
  \bibinfo {author} {\bibfnamefont {A.~A.}\ \bibnamefont {Coelho}}, \bibinfo
  {author} {\bibfnamefont {A.~K.}\ \bibnamefont {Nigam}}, \bibinfo {author}
  {\bibfnamefont {D.~D.}\ \bibnamefont {Johnson}}, \bibinfo {author}
  {\bibfnamefont {A.}~\bibnamefont {Alam}},\ and\ \bibinfo {author}
  {\bibfnamefont {K.~G.}\ \bibnamefont {Suresh}},\ }\href@noop {} {\bibfield
  {journal} {\bibinfo  {journal} {Phys. Rev. B}\ }\textbf {\bibinfo {volume}
  {92}},\ \bibinfo {pages} {045201} (\bibinfo {year}
  {2015}{\natexlab{b}})}\BibitemShut {NoStop}%
\bibitem [{\citenamefont {Bainsla}\ and\ \citenamefont
  {Suresh}(2016)}]{bainsla2016equiatomic}%
  \BibitemOpen
  \bibfield  {author} {\bibinfo {author} {\bibfnamefont {L.}~\bibnamefont
  {Bainsla}}\ and\ \bibinfo {author} {\bibfnamefont {K.}~\bibnamefont
  {Suresh}},\ }\href@noop {} {\bibfield  {journal} {\bibinfo  {journal} {Appl.
  Phys. Rev.}\ }\textbf {\bibinfo {volume} {3}},\ \bibinfo {pages} {031101}
  (\bibinfo {year} {2016})}\BibitemShut {NoStop}%
\bibitem [{\citenamefont {Alijani}\ \emph
  {et~al.}(2011{\natexlab{a}})\citenamefont {Alijani}, \citenamefont {Ouardi},
  \citenamefont {Fecher}, \citenamefont {Winterlik}, \citenamefont {Naghavi},
  \citenamefont {Kozina}, \citenamefont {Stryganyuk}, \citenamefont {Felser},
  \citenamefont {Ikenaga}, \citenamefont {Yamashita} \emph
  {et~al.}}]{alijani2011electronic}%
  \BibitemOpen
  \bibfield  {author} {\bibinfo {author} {\bibfnamefont {V.}~\bibnamefont
  {Alijani}}, \bibinfo {author} {\bibfnamefont {S.}~\bibnamefont {Ouardi}},
  \bibinfo {author} {\bibfnamefont {G.~H.}\ \bibnamefont {Fecher}}, \bibinfo
  {author} {\bibfnamefont {J.}~\bibnamefont {Winterlik}}, \bibinfo {author}
  {\bibfnamefont {S.~S.}\ \bibnamefont {Naghavi}}, \bibinfo {author}
  {\bibfnamefont {X.}~\bibnamefont {Kozina}}, \bibinfo {author} {\bibfnamefont
  {G.}~\bibnamefont {Stryganyuk}}, \bibinfo {author} {\bibfnamefont
  {C.}~\bibnamefont {Felser}}, \bibinfo {author} {\bibfnamefont
  {E.}~\bibnamefont {Ikenaga}}, \bibinfo {author} {\bibfnamefont
  {Y.}~\bibnamefont {Yamashita}}, \emph {et~al.},\ }\href@noop {} {\bibfield
  {journal} {\bibinfo  {journal} {Phys. Rev. B}\ }\textbf {\bibinfo {volume}
  {84}},\ \bibinfo {pages} {224416} (\bibinfo {year}
  {2011}{\natexlab{a}})}\BibitemShut {NoStop}%
\bibitem [{\citenamefont {Alijani}\ \emph
  {et~al.}(2011{\natexlab{b}})\citenamefont {Alijani}, \citenamefont
  {Winterlik}, \citenamefont {Fecher}, \citenamefont {Naghavi},\ and\
  \citenamefont {Felser}}]{alijani2011quaternary}%
  \BibitemOpen
  \bibfield  {author} {\bibinfo {author} {\bibfnamefont {V.}~\bibnamefont
  {Alijani}}, \bibinfo {author} {\bibfnamefont {J.}~\bibnamefont {Winterlik}},
  \bibinfo {author} {\bibfnamefont {G.~H.}\ \bibnamefont {Fecher}}, \bibinfo
  {author} {\bibfnamefont {S.~S.}\ \bibnamefont {Naghavi}},\ and\ \bibinfo
  {author} {\bibfnamefont {C.}~\bibnamefont {Felser}},\ }\href@noop {}
  {\bibfield  {journal} {\bibinfo  {journal} {Phys. Rev. B}\ }\textbf {\bibinfo
  {volume} {83}},\ \bibinfo {pages} {184428} (\bibinfo {year}
  {2011}{\natexlab{b}})}\BibitemShut {NoStop}%
\bibitem [{\citenamefont {Venkateswara}\ \emph {et~al.}(2019)\citenamefont
  {Venkateswara}, \citenamefont {Samatham}, \citenamefont {Babu}, \citenamefont
  {Suresh},\ and\ \citenamefont {Alam}}]{venkateswara2019coexistence}%
  \BibitemOpen
  \bibfield  {author} {\bibinfo {author} {\bibfnamefont {Y.}~\bibnamefont
  {Venkateswara}}, \bibinfo {author} {\bibfnamefont {S.~S.}\ \bibnamefont
  {Samatham}}, \bibinfo {author} {\bibfnamefont {P.~D.}\ \bibnamefont {Babu}},
  \bibinfo {author} {\bibfnamefont {K.~G.}\ \bibnamefont {Suresh}},\ and\
  \bibinfo {author} {\bibfnamefont {A.}~\bibnamefont {Alam}},\ }\href@noop {}
  {\bibfield  {journal} {\bibinfo  {journal} {Phys. Rev. B}\ }\textbf {\bibinfo
  {volume} {100}},\ \bibinfo {pages} {180404} (\bibinfo {year}
  {2019})}\BibitemShut {NoStop}%
\bibitem [{\citenamefont {Wang}(2008)}]{wang2008proposal}%
  \BibitemOpen
  \bibfield  {author} {\bibinfo {author} {\bibfnamefont {X.~L.}\ \bibnamefont
  {Wang}},\ }\href@noop {} {\bibfield  {journal} {\bibinfo  {journal} {Phys.
  Rev. Lett.}\ }\textbf {\bibinfo {volume} {100}},\ \bibinfo {pages} {156404}
  (\bibinfo {year} {2008})}\BibitemShut {NoStop}%
\bibitem [{\citenamefont {Kundu}\ \emph {et~al.}(2017)\citenamefont {Kundu},
  \citenamefont {Ghosh}, \citenamefont {Banerjee}, \citenamefont {Ghosh},\ and\
  \citenamefont {Sanyal}}]{kundu2017new}%
  \BibitemOpen
  \bibfield  {author} {\bibinfo {author} {\bibfnamefont {A.}~\bibnamefont
  {Kundu}}, \bibinfo {author} {\bibfnamefont {S.}~\bibnamefont {Ghosh}},
  \bibinfo {author} {\bibfnamefont {R.}~\bibnamefont {Banerjee}}, \bibinfo
  {author} {\bibfnamefont {S.}~\bibnamefont {Ghosh}},\ and\ \bibinfo {author}
  {\bibfnamefont {B.}~\bibnamefont {Sanyal}},\ }\href@noop {} {\bibfield
  {journal} {\bibinfo  {journal} {Sci. Rep.}\ }\textbf {\bibinfo {volume}
  {7}},\ \bibinfo {pages} {1} (\bibinfo {year} {2017})}\BibitemShut {NoStop}%
\bibitem [{\citenamefont {Gao}\ \emph {et~al.}(2019)\citenamefont {Gao},
  \citenamefont {Opahle},\ and\ \citenamefont {Zhang}}]{gao2019high}%
  \BibitemOpen
  \bibfield  {author} {\bibinfo {author} {\bibfnamefont {Q.}~\bibnamefont
  {Gao}}, \bibinfo {author} {\bibfnamefont {I.}~\bibnamefont {Opahle}},\ and\
  \bibinfo {author} {\bibfnamefont {H.}~\bibnamefont {Zhang}},\ }\href@noop {}
  {\bibfield  {journal} {\bibinfo  {journal} {Phys. Rev. Mater.}\ }\textbf
  {\bibinfo {volume} {3}},\ \bibinfo {pages} {024410} (\bibinfo {year}
  {2019})}\BibitemShut {NoStop}%
\bibitem [{\citenamefont {{\"O}zdo{\u{g}}an}\ \emph {et~al.}(2013)\citenamefont
  {{\"O}zdo{\u{g}}an}, \citenamefont {{\c{S}}a{\c{s}}{\i}o{\u{g}}lu},\ and\
  \citenamefont {Galanakis}}]{ozdougan2013slater}%
  \BibitemOpen
  \bibfield  {author} {\bibinfo {author} {\bibfnamefont {K.}~\bibnamefont
  {{\"O}zdo{\u{g}}an}}, \bibinfo {author} {\bibfnamefont {E.}~\bibnamefont
  {{\c{S}}a{\c{s}}{\i}o{\u{g}}lu}},\ and\ \bibinfo {author} {\bibfnamefont
  {I.}~\bibnamefont {Galanakis}},\ }\href@noop {} {\bibfield  {journal}
  {\bibinfo  {journal} {Journal of Applied Physics}\ }\textbf {\bibinfo
  {volume} {113}},\ \bibinfo {pages} {193903} (\bibinfo {year}
  {2013})}\BibitemShut {NoStop}%
\bibitem [{\citenamefont {Herran}\ \emph {et~al.}(2017)\citenamefont {Herran},
  \citenamefont {Dalal}, \citenamefont {Gray}, \citenamefont {Kharel},\ and\
  \citenamefont {Lukashev}}]{herran2017atomic}%
  \BibitemOpen
  \bibfield  {author} {\bibinfo {author} {\bibfnamefont {J.}~\bibnamefont
  {Herran}}, \bibinfo {author} {\bibfnamefont {R.}~\bibnamefont {Dalal}},
  \bibinfo {author} {\bibfnamefont {P.}~\bibnamefont {Gray}}, \bibinfo {author}
  {\bibfnamefont {P.}~\bibnamefont {Kharel}},\ and\ \bibinfo {author}
  {\bibfnamefont {P.~V.}\ \bibnamefont {Lukashev}},\ }\href@noop {} {\bibfield
  {journal} {\bibinfo  {journal} {Journal of Applied Physics}\ }\textbf
  {\bibinfo {volume} {122}},\ \bibinfo {pages} {153904} (\bibinfo {year}
  {2017})}\BibitemShut {NoStop}%
\bibitem [{\citenamefont {Kharel}\ \emph {et~al.}(2017)\citenamefont {Kharel},
  \citenamefont {Herran}, \citenamefont {Lukashev}, \citenamefont {Jin},
  \citenamefont {Waybright}, \citenamefont {Gilbert}, \citenamefont {Staten},
  \citenamefont {Gray}, \citenamefont {Valloppilly}, \citenamefont {Huh} \emph
  {et~al.}}]{kharel2017effect}%
  \BibitemOpen
  \bibfield  {author} {\bibinfo {author} {\bibfnamefont {P.}~\bibnamefont
  {Kharel}}, \bibinfo {author} {\bibfnamefont {J.}~\bibnamefont {Herran}},
  \bibinfo {author} {\bibfnamefont {P.}~\bibnamefont {Lukashev}}, \bibinfo
  {author} {\bibfnamefont {Y.}~\bibnamefont {Jin}}, \bibinfo {author}
  {\bibfnamefont {J.}~\bibnamefont {Waybright}}, \bibinfo {author}
  {\bibfnamefont {S.}~\bibnamefont {Gilbert}}, \bibinfo {author} {\bibfnamefont
  {B.}~\bibnamefont {Staten}}, \bibinfo {author} {\bibfnamefont
  {P.}~\bibnamefont {Gray}}, \bibinfo {author} {\bibfnamefont {S.}~\bibnamefont
  {Valloppilly}}, \bibinfo {author} {\bibfnamefont {Y.}~\bibnamefont {Huh}},
  \emph {et~al.},\ }\href@noop {} {\bibfield  {journal} {\bibinfo  {journal}
  {AIP Advances}\ }\textbf {\bibinfo {volume} {7}},\ \bibinfo {pages} {056402}
  (\bibinfo {year} {2017})}\BibitemShut {NoStop}%
\bibitem [{\citenamefont {Mukadam}\ \emph {et~al.}(2016)\citenamefont
  {Mukadam}, \citenamefont {Roy}, \citenamefont {Meena}, \citenamefont
  {Bhatt},\ and\ \citenamefont {Yusuf}}]{mukadam2016quantification}%
  \BibitemOpen
  \bibfield  {author} {\bibinfo {author} {\bibfnamefont {M.~D.}\ \bibnamefont
  {Mukadam}}, \bibinfo {author} {\bibfnamefont {S.}~\bibnamefont {Roy}},
  \bibinfo {author} {\bibfnamefont {S.~S.}\ \bibnamefont {Meena}}, \bibinfo
  {author} {\bibfnamefont {P.}~\bibnamefont {Bhatt}},\ and\ \bibinfo {author}
  {\bibfnamefont {S.~M.}\ \bibnamefont {Yusuf}},\ }\href@noop {} {\bibfield
  {journal} {\bibinfo  {journal} {Phys. Rev. B}\ }\textbf {\bibinfo {volume}
  {94}},\ \bibinfo {pages} {214423} (\bibinfo {year} {2016})}\BibitemShut
  {NoStop}%
\bibitem [{\citenamefont {Avdeev}\ and\ \citenamefont
  {Hester}(2018)}]{avdeev2018echidna}%
  \BibitemOpen
  \bibfield  {author} {\bibinfo {author} {\bibfnamefont {M.}~\bibnamefont
  {Avdeev}}\ and\ \bibinfo {author} {\bibfnamefont {J.~R.}\ \bibnamefont
  {Hester}},\ }\href@noop {} {\bibfield  {journal} {\bibinfo  {journal}
  {Journal of Applied Crystallography}\ }\textbf {\bibinfo {volume} {51}},\
  \bibinfo {pages} {1597} (\bibinfo {year} {2018})}\BibitemShut {NoStop}%
\bibitem [{\citenamefont
  {Rodr{\'\i}guez-Carvajal}(1993)}]{rodriguez1993recent}%
  \BibitemOpen
  \bibfield  {author} {\bibinfo {author} {\bibfnamefont {J.}~\bibnamefont
  {Rodr{\'\i}guez-Carvajal}},\ }\href@noop {} {\bibfield  {journal} {\bibinfo
  {journal} {Phys. B: Condens. Matter}\ }\textbf {\bibinfo {volume} {192}},\
  \bibinfo {pages} {55} (\bibinfo {year} {1993})}\BibitemShut {NoStop}%
\bibitem [{\citenamefont {Bl{\"o}chl}(1994)}]{blochl1994projector}%
  \BibitemOpen
  \bibfield  {author} {\bibinfo {author} {\bibfnamefont {P.~E.}\ \bibnamefont
  {Bl{\"o}chl}},\ }\href@noop {} {\bibfield  {journal} {\bibinfo  {journal}
  {Phys. Rev. B}\ }\textbf {\bibinfo {volume} {50}},\ \bibinfo {pages} {17953}
  (\bibinfo {year} {1994})}\BibitemShut {NoStop}%
\bibitem [{\citenamefont {Kresse}\ and\ \citenamefont
  {Hafner}(1993)}]{kresse1993ab}%
  \BibitemOpen
  \bibfield  {author} {\bibinfo {author} {\bibfnamefont {G.}~\bibnamefont
  {Kresse}}\ and\ \bibinfo {author} {\bibfnamefont {J.}~\bibnamefont
  {Hafner}},\ }\href@noop {} {\bibfield  {journal} {\bibinfo  {journal} {Phys.
  Rev. B}\ }\textbf {\bibinfo {volume} {48}},\ \bibinfo {pages} {13115}
  (\bibinfo {year} {1993})}\BibitemShut {NoStop}%
\bibitem [{\citenamefont {Kresse}\ and\ \citenamefont
  {Hafner}(1994)}]{kresse1994norm}%
  \BibitemOpen
  \bibfield  {author} {\bibinfo {author} {\bibfnamefont {G.}~\bibnamefont
  {Kresse}}\ and\ \bibinfo {author} {\bibfnamefont {J.}~\bibnamefont
  {Hafner}},\ }\href@noop {} {\bibfield  {journal} {\bibinfo  {journal} {J.
  Phys. Condens. Matter}\ }\textbf {\bibinfo {volume} {6}},\ \bibinfo {pages}
  {8245} (\bibinfo {year} {1994})}\BibitemShut {NoStop}%
\bibitem [{\citenamefont {Perdew}\ \emph {et~al.}(1996)\citenamefont {Perdew},
  \citenamefont {Burke},\ and\ \citenamefont
  {Ernzerhof}}]{perdew1996generalized}%
  \BibitemOpen
  \bibfield  {author} {\bibinfo {author} {\bibfnamefont {J.~P.}\ \bibnamefont
  {Perdew}}, \bibinfo {author} {\bibfnamefont {K.}~\bibnamefont {Burke}},\ and\
  \bibinfo {author} {\bibfnamefont {M.}~\bibnamefont {Ernzerhof}},\ }\href@noop
  {} {\bibfield  {journal} {\bibinfo  {journal} {Phys. Rev. Lett.}\ }\textbf
  {\bibinfo {volume} {77}},\ \bibinfo {pages} {3865} (\bibinfo {year}
  {1996})}\BibitemShut {NoStop}%
\bibitem [{\citenamefont {Bl{\"o}chl}\ \emph {et~al.}(1994)\citenamefont
  {Bl{\"o}chl}, \citenamefont {Jepsen},\ and\ \citenamefont
  {Andersen}}]{blochl1994improved}%
  \BibitemOpen
  \bibfield  {author} {\bibinfo {author} {\bibfnamefont {P.~E.}\ \bibnamefont
  {Bl{\"o}chl}}, \bibinfo {author} {\bibfnamefont {O.}~\bibnamefont {Jepsen}},\
  and\ \bibinfo {author} {\bibfnamefont {O.~K.}\ \bibnamefont {Andersen}},\
  }\href@noop {} {\bibfield  {journal} {\bibinfo  {journal} {Phys. Rev. B}\
  }\textbf {\bibinfo {volume} {49}},\ \bibinfo {pages} {16223} (\bibinfo {year}
  {1994})}\BibitemShut {NoStop}%
\bibitem [{\citenamefont {Zunger}\ \emph {et~al.}(1990)\citenamefont {Zunger},
  \citenamefont {Wei}, \citenamefont {Ferreira},\ and\ \citenamefont
  {Bernard}}]{zunger1990special}%
  \BibitemOpen
  \bibfield  {author} {\bibinfo {author} {\bibfnamefont {A.}~\bibnamefont
  {Zunger}}, \bibinfo {author} {\bibfnamefont {S.-H.}\ \bibnamefont {Wei}},
  \bibinfo {author} {\bibfnamefont {L.~G.}\ \bibnamefont {Ferreira}},\ and\
  \bibinfo {author} {\bibfnamefont {J.~E.}\ \bibnamefont {Bernard}},\
  }\href@noop {} {\bibfield  {journal} {\bibinfo  {journal} {Phys. Rev. Lett.}\
  }\textbf {\bibinfo {volume} {65}},\ \bibinfo {pages} {353} (\bibinfo {year}
  {1990})}\BibitemShut {NoStop}%
\bibitem [{\citenamefont {Sanchez}\ \emph {et~al.}(1984)\citenamefont
  {Sanchez}, \citenamefont {Ducastelle},\ and\ \citenamefont
  {Gratias}}]{sanchez1984generalized}%
  \BibitemOpen
  \bibfield  {author} {\bibinfo {author} {\bibfnamefont {J.~M.}\ \bibnamefont
  {Sanchez}}, \bibinfo {author} {\bibfnamefont {F.}~\bibnamefont
  {Ducastelle}},\ and\ \bibinfo {author} {\bibfnamefont {D.}~\bibnamefont
  {Gratias}},\ }\href@noop {} {\bibfield  {journal} {\bibinfo  {journal} {Phys.
  A: Stat. Mech. Appl.}\ }\textbf {\bibinfo {volume} {128}},\ \bibinfo {pages}
  {334} (\bibinfo {year} {1984})}\BibitemShut {NoStop}%
\bibitem [{\citenamefont {Van De~Walle}(2009)}]{van2009multicomponent}%
  \BibitemOpen
  \bibfield  {author} {\bibinfo {author} {\bibfnamefont {A.}~\bibnamefont {Van
  De~Walle}},\ }\href@noop {} {\bibfield  {journal} {\bibinfo  {journal}
  {Calphad}\ }\textbf {\bibinfo {volume} {33}},\ \bibinfo {pages} {266}
  (\bibinfo {year} {2009})}\BibitemShut {NoStop}%
\bibitem [{\citenamefont {Van~de Walle}\ \emph {et~al.}(2013)\citenamefont
  {Van~de Walle}, \citenamefont {Tiwary}, \citenamefont {De~Jong},
  \citenamefont {Olmsted}, \citenamefont {Asta}, \citenamefont {Dick},
  \citenamefont {Shin}, \citenamefont {Wang}, \citenamefont {Chen},\ and\
  \citenamefont {Liu}}]{van2013efficient}%
  \BibitemOpen
  \bibfield  {author} {\bibinfo {author} {\bibfnamefont {A.}~\bibnamefont
  {Van~de Walle}}, \bibinfo {author} {\bibfnamefont {P.}~\bibnamefont
  {Tiwary}}, \bibinfo {author} {\bibfnamefont {M.}~\bibnamefont {De~Jong}},
  \bibinfo {author} {\bibfnamefont {D.~L.}\ \bibnamefont {Olmsted}}, \bibinfo
  {author} {\bibfnamefont {M.}~\bibnamefont {Asta}}, \bibinfo {author}
  {\bibfnamefont {A.}~\bibnamefont {Dick}}, \bibinfo {author} {\bibfnamefont
  {D.}~\bibnamefont {Shin}}, \bibinfo {author} {\bibfnamefont {Y.}~\bibnamefont
  {Wang}}, \bibinfo {author} {\bibfnamefont {L.-Q.}\ \bibnamefont {Chen}},\
  and\ \bibinfo {author} {\bibfnamefont {Z.-K.}\ \bibnamefont {Liu}},\
  }\href@noop {} {\bibfield  {journal} {\bibinfo  {journal} {Calphad}\ }\textbf
  {\bibinfo {volume} {42}},\ \bibinfo {pages} {13} (\bibinfo {year}
  {2013})}\BibitemShut {NoStop}%
\bibitem [{\citenamefont {Diack-Rasselio}\ \emph {et~al.}(2022)\citenamefont
  {Diack-Rasselio}, \citenamefont {Rouleau}, \citenamefont {Coulomb},
  \citenamefont {Georgeton}, \citenamefont {Beaudhuin}, \citenamefont
  {Crivello},\ and\ \citenamefont {Alleno}}]{diack2022influence}%
  \BibitemOpen
  \bibfield  {author} {\bibinfo {author} {\bibfnamefont {A.}~\bibnamefont
  {Diack-Rasselio}}, \bibinfo {author} {\bibfnamefont {O.}~\bibnamefont
  {Rouleau}}, \bibinfo {author} {\bibfnamefont {L.}~\bibnamefont {Coulomb}},
  \bibinfo {author} {\bibfnamefont {L.}~\bibnamefont {Georgeton}}, \bibinfo
  {author} {\bibfnamefont {M.}~\bibnamefont {Beaudhuin}}, \bibinfo {author}
  {\bibfnamefont {J.-C.}\ \bibnamefont {Crivello}},\ and\ \bibinfo {author}
  {\bibfnamefont {E.}~\bibnamefont {Alleno}},\ }\href@noop {} {\bibfield
  {journal} {\bibinfo  {journal} {Journal of Alloys and Compounds}\ }\textbf
  {\bibinfo {volume} {920}},\ \bibinfo {pages} {166037} (\bibinfo {year}
  {2022})}\BibitemShut {NoStop}%
\bibitem [{\citenamefont {Galanakis}\ \emph {et~al.}(2002)\citenamefont
  {Galanakis}, \citenamefont {Dederichs},\ and\ \citenamefont
  {Papanikolaou}}]{galanakis2002slater}%
  \BibitemOpen
  \bibfield  {author} {\bibinfo {author} {\bibfnamefont {I.}~\bibnamefont
  {Galanakis}}, \bibinfo {author} {\bibfnamefont {P.~H.}\ \bibnamefont
  {Dederichs}},\ and\ \bibinfo {author} {\bibfnamefont {N.}~\bibnamefont
  {Papanikolaou}},\ }\href@noop {} {\bibfield  {journal} {\bibinfo  {journal}
  {Phys. Rev. B}\ }\textbf {\bibinfo {volume} {66}},\ \bibinfo {pages} {174429}
  (\bibinfo {year} {2002})}\BibitemShut {NoStop}%
\bibitem [{\citenamefont {Galanakis}\ \emph {et~al.}(2007)\citenamefont
  {Galanakis}, \citenamefont {{\"O}zdo{\u{g}}an}, \citenamefont
  {{\c{S}}a{\c{s}}{\i}o{\u{g}}lu},\ and\ \citenamefont
  {Akta{\c{s}}}}]{galanakis2007doping}%
  \BibitemOpen
  \bibfield  {author} {\bibinfo {author} {\bibfnamefont {I.}~\bibnamefont
  {Galanakis}}, \bibinfo {author} {\bibfnamefont {K.}~\bibnamefont
  {{\"O}zdo{\u{g}}an}}, \bibinfo {author} {\bibfnamefont {E.}~\bibnamefont
  {{\c{S}}a{\c{s}}{\i}o{\u{g}}lu}},\ and\ \bibinfo {author} {\bibfnamefont
  {B.}~\bibnamefont {Akta{\c{s}}}},\ }\href@noop {} {\bibfield  {journal}
  {\bibinfo  {journal} {Physical Review B}\ }\textbf {\bibinfo {volume} {75}},\
  \bibinfo {pages} {092407} (\bibinfo {year} {2007})}\BibitemShut {NoStop}%
\bibitem [{\citenamefont {Samanta}\ \emph {et~al.}(2018)\citenamefont
  {Samanta}, \citenamefont {Bhobe}, \citenamefont {Das}, \citenamefont
  {Kumar},\ and\ \citenamefont {Nigam}}]{samanta2018reentrant}%
  \BibitemOpen
  \bibfield  {author} {\bibinfo {author} {\bibfnamefont {T.}~\bibnamefont
  {Samanta}}, \bibinfo {author} {\bibfnamefont {P.~A.}\ \bibnamefont {Bhobe}},
  \bibinfo {author} {\bibfnamefont {A.}~\bibnamefont {Das}}, \bibinfo {author}
  {\bibfnamefont {A.}~\bibnamefont {Kumar}},\ and\ \bibinfo {author}
  {\bibfnamefont {A.}~\bibnamefont {Nigam}},\ }\href@noop {} {\bibfield
  {journal} {\bibinfo  {journal} {Phys. Rev. B}\ }\textbf {\bibinfo {volume}
  {97}},\ \bibinfo {pages} {184421} (\bibinfo {year} {2018})}\BibitemShut
  {NoStop}%
\bibitem [{\citenamefont {Raphael}\ \emph {et~al.}(2002)\citenamefont
  {Raphael}, \citenamefont {Ravel}, \citenamefont {Huang}, \citenamefont
  {Willard}, \citenamefont {Cheng}, \citenamefont {Das}, \citenamefont
  {Stroud}, \citenamefont {Bussmann}, \citenamefont {Claassen},\ and\
  \citenamefont {Harris}}]{raphael2002presence}%
  \BibitemOpen
  \bibfield  {author} {\bibinfo {author} {\bibfnamefont {M.}~\bibnamefont
  {Raphael}}, \bibinfo {author} {\bibfnamefont {B.}~\bibnamefont {Ravel}},
  \bibinfo {author} {\bibfnamefont {Q.}~\bibnamefont {Huang}}, \bibinfo
  {author} {\bibfnamefont {M.~A.}\ \bibnamefont {Willard}}, \bibinfo {author}
  {\bibfnamefont {S.~F.}\ \bibnamefont {Cheng}}, \bibinfo {author}
  {\bibfnamefont {B.~N.}\ \bibnamefont {Das}}, \bibinfo {author} {\bibfnamefont
  {R.~M.}\ \bibnamefont {Stroud}}, \bibinfo {author} {\bibfnamefont {K.~M.}\
  \bibnamefont {Bussmann}}, \bibinfo {author} {\bibfnamefont {J.~H.}\
  \bibnamefont {Claassen}},\ and\ \bibinfo {author} {\bibfnamefont {V.~G.}\
  \bibnamefont {Harris}},\ }\href@noop {} {\bibfield  {journal} {\bibinfo
  {journal} {Phys. Rev. B}\ }\textbf {\bibinfo {volume} {66}},\ \bibinfo
  {pages} {104429} (\bibinfo {year} {2002})}\BibitemShut {NoStop}%
\bibitem [{\citenamefont {Miura}\ \emph {et~al.}(2004)\citenamefont {Miura},
  \citenamefont {Nagao},\ and\ \citenamefont {Shirai}}]{miura2004atomic}%
  \BibitemOpen
  \bibfield  {author} {\bibinfo {author} {\bibfnamefont {Y.}~\bibnamefont
  {Miura}}, \bibinfo {author} {\bibfnamefont {K.}~\bibnamefont {Nagao}},\ and\
  \bibinfo {author} {\bibfnamefont {M.}~\bibnamefont {Shirai}},\ }\href@noop {}
  {\bibfield  {journal} {\bibinfo  {journal} {Phys. Rev. B}\ }\textbf {\bibinfo
  {volume} {69}},\ \bibinfo {pages} {144413} (\bibinfo {year}
  {2004})}\BibitemShut {NoStop}%
\bibitem [{\citenamefont {Webster}\ and\ \citenamefont
  {Ziebeck}(1973)}]{webster1973magnetic}%
  \BibitemOpen
  \bibfield  {author} {\bibinfo {author} {\bibfnamefont {P.~J.}\ \bibnamefont
  {Webster}}\ and\ \bibinfo {author} {\bibfnamefont {K.~R.~A.}\ \bibnamefont
  {Ziebeck}},\ }\href@noop {} {\bibfield  {journal} {\bibinfo  {journal} {J.
  Phys. Chem. Solids}\ }\textbf {\bibinfo {volume} {34}},\ \bibinfo {pages}
  {1647} (\bibinfo {year} {1973})}\BibitemShut {NoStop}%
\bibitem [{\citenamefont {Venkateswara}\ \emph {et~al.}(2015)\citenamefont
  {Venkateswara}, \citenamefont {Gupta}, \citenamefont {Varma}, \citenamefont
  {Singh}, \citenamefont {Suresh},\ and\ \citenamefont
  {Alam}}]{venkateswara2015electronic}%
  \BibitemOpen
  \bibfield  {author} {\bibinfo {author} {\bibfnamefont {Y.}~\bibnamefont
  {Venkateswara}}, \bibinfo {author} {\bibfnamefont {S.}~\bibnamefont {Gupta}},
  \bibinfo {author} {\bibfnamefont {M.~R.}\ \bibnamefont {Varma}}, \bibinfo
  {author} {\bibfnamefont {P.}~\bibnamefont {Singh}}, \bibinfo {author}
  {\bibfnamefont {K.~G.}\ \bibnamefont {Suresh}},\ and\ \bibinfo {author}
  {\bibfnamefont {A.}~\bibnamefont {Alam}},\ }\href@noop {} {\bibfield
  {journal} {\bibinfo  {journal} {Phys. Rev. B}\ }\textbf {\bibinfo {volume}
  {92}},\ \bibinfo {pages} {224413} (\bibinfo {year} {2015})}\BibitemShut
  {NoStop}%
\bibitem [{\citenamefont {Balke}\ \emph {et~al.}(2007)\citenamefont {Balke},
  \citenamefont {Wurmehl}, \citenamefont {Fecher}, \citenamefont {Felser},
  \citenamefont {Alves}, \citenamefont {Bernardi},\ and\ \citenamefont
  {Morais}}]{balke2007structural}%
  \BibitemOpen
  \bibfield  {author} {\bibinfo {author} {\bibfnamefont {B.}~\bibnamefont
  {Balke}}, \bibinfo {author} {\bibfnamefont {S.}~\bibnamefont {Wurmehl}},
  \bibinfo {author} {\bibfnamefont {G.~H.}\ \bibnamefont {Fecher}}, \bibinfo
  {author} {\bibfnamefont {C.}~\bibnamefont {Felser}}, \bibinfo {author}
  {\bibfnamefont {M.~C.~M.}\ \bibnamefont {Alves}}, \bibinfo {author}
  {\bibfnamefont {F.}~\bibnamefont {Bernardi}},\ and\ \bibinfo {author}
  {\bibfnamefont {J.}~\bibnamefont {Morais}},\ }\href@noop {} {\bibfield
  {journal} {\bibinfo  {journal} {Appl. Phys. Lett.}\ }\textbf {\bibinfo
  {volume} {90}},\ \bibinfo {pages} {172501} (\bibinfo {year}
  {2007})}\BibitemShut {NoStop}%
\bibitem [{\citenamefont {Bainsla}\ \emph
  {et~al.}(2015{\natexlab{c}})\citenamefont {Bainsla}, \citenamefont {Yadav},
  \citenamefont {Venkateswara}, \citenamefont {Jha}, \citenamefont
  {Bhattacharyya},\ and\ \citenamefont {Suresh}}]{bainsla2015local}%
  \BibitemOpen
  \bibfield  {author} {\bibinfo {author} {\bibfnamefont {L.}~\bibnamefont
  {Bainsla}}, \bibinfo {author} {\bibfnamefont {A.~K.}\ \bibnamefont {Yadav}},
  \bibinfo {author} {\bibfnamefont {Y.}~\bibnamefont {Venkateswara}}, \bibinfo
  {author} {\bibfnamefont {S.~N.}\ \bibnamefont {Jha}}, \bibinfo {author}
  {\bibfnamefont {D.}~\bibnamefont {Bhattacharyya}},\ and\ \bibinfo {author}
  {\bibfnamefont {K.~G.}\ \bibnamefont {Suresh}},\ }\href@noop {} {\bibfield
  {journal} {\bibinfo  {journal} {J. Alloys Compd.}\ }\textbf {\bibinfo
  {volume} {651}},\ \bibinfo {pages} {509} (\bibinfo {year}
  {2015}{\natexlab{c}})}\BibitemShut {NoStop}%
\bibitem [{\citenamefont {Ravel}\ \emph {et~al.}(2002)\citenamefont {Ravel},
  \citenamefont {Raphael}, \citenamefont {Harris},\ and\ \citenamefont
  {Huang}}]{ravel2002exafs}%
  \BibitemOpen
  \bibfield  {author} {\bibinfo {author} {\bibfnamefont {B.}~\bibnamefont
  {Ravel}}, \bibinfo {author} {\bibfnamefont {M.~P.}\ \bibnamefont {Raphael}},
  \bibinfo {author} {\bibfnamefont {V.~G.}\ \bibnamefont {Harris}},\ and\
  \bibinfo {author} {\bibfnamefont {Q.}~\bibnamefont {Huang}},\ }\href@noop {}
  {\bibfield  {journal} {\bibinfo  {journal} {Phys. Rev. B}\ }\textbf {\bibinfo
  {volume} {65}},\ \bibinfo {pages} {184431} (\bibinfo {year}
  {2002})}\BibitemShut {NoStop}%
\bibitem [{\citenamefont {Koningsberger}\ and\ \citenamefont
  {Prins}(1987)}]{koningsberger1987x}%
  \BibitemOpen
  \bibfield  {author} {\bibinfo {author} {\bibfnamefont {D.~C.}\ \bibnamefont
  {Koningsberger}}\ and\ \bibinfo {author} {\bibfnamefont {R.}~\bibnamefont
  {Prins}},\ }\href@noop {} {\  (\bibinfo {year} {1987})}\BibitemShut {NoStop}%
\bibitem [{\citenamefont {Ravel}\ and\ \citenamefont
  {Newville}(2005)}]{ravel2005athena}%
  \BibitemOpen
  \bibfield  {author} {\bibinfo {author} {\bibfnamefont {B.}~\bibnamefont
  {Ravel}}\ and\ \bibinfo {author} {\bibfnamefont {M.}~\bibnamefont
  {Newville}},\ }\href@noop {} {\bibfield  {journal} {\bibinfo  {journal} {J.
  Synchrotron Radiat.}\ }\textbf {\bibinfo {volume} {12}},\ \bibinfo {pages}
  {537} (\bibinfo {year} {2005})}\BibitemShut {NoStop}%
\bibitem [{\citenamefont {Galanakis}\ \emph {et~al.}(2006)\citenamefont
  {Galanakis}, \citenamefont {Mavropoulos},\ and\ \citenamefont
  {Dederichs}}]{galanakis2006electronic}%
  \BibitemOpen
  \bibfield  {author} {\bibinfo {author} {\bibfnamefont {I.}~\bibnamefont
  {Galanakis}}, \bibinfo {author} {\bibfnamefont {P.}~\bibnamefont
  {Mavropoulos}},\ and\ \bibinfo {author} {\bibfnamefont {P.~H.}\ \bibnamefont
  {Dederichs}},\ }\href@noop {} {\bibfield  {journal} {\bibinfo  {journal} {J.
  Phys. D}\ }\textbf {\bibinfo {volume} {39}},\ \bibinfo {pages} {765}
  (\bibinfo {year} {2006})}\BibitemShut {NoStop}%
\bibitem [{\citenamefont {Rani}\ \emph {et~al.}(2017)\citenamefont {Rani},
  \citenamefont {Suresh}, \citenamefont {Yadav}, \citenamefont {Jha},
  \citenamefont {Bhattacharyya}, \citenamefont {Varma},\ and\ \citenamefont
  {Alam}}]{rani2017structural}%
  \BibitemOpen
  \bibfield  {author} {\bibinfo {author} {\bibfnamefont {D.}~\bibnamefont
  {Rani}}, \bibinfo {author} {\bibfnamefont {K.~G.}\ \bibnamefont {Suresh}},
  \bibinfo {author} {\bibfnamefont {A.~K.}\ \bibnamefont {Yadav}}, \bibinfo
  {author} {\bibfnamefont {S.~N.}\ \bibnamefont {Jha}}, \bibinfo {author}
  {\bibfnamefont {D.}~\bibnamefont {Bhattacharyya}}, \bibinfo {author}
  {\bibfnamefont {M.~R.}\ \bibnamefont {Varma}},\ and\ \bibinfo {author}
  {\bibfnamefont {A.}~\bibnamefont {Alam}},\ }\href@noop {} {\bibfield
  {journal} {\bibinfo  {journal} {Phys. Rev. B}\ }\textbf {\bibinfo {volume}
  {96}},\ \bibinfo {pages} {184404} (\bibinfo {year} {2017})}\BibitemShut
  {NoStop}%
\bibitem [{\citenamefont {B{\oe}uf}\ \emph {et~al.}(2006)\citenamefont
  {B{\oe}uf}, \citenamefont {Pfleiderer},\ and\ \citenamefont
  {Fai{\ss}t}}]{boeuf2006low}%
  \BibitemOpen
  \bibfield  {author} {\bibinfo {author} {\bibfnamefont {J.}~\bibnamefont
  {B{\oe}uf}}, \bibinfo {author} {\bibfnamefont {C.}~\bibnamefont
  {Pfleiderer}},\ and\ \bibinfo {author} {\bibfnamefont {A.}~\bibnamefont
  {Fai{\ss}t}},\ }\href@noop {} {\bibfield  {journal} {\bibinfo  {journal}
  {Phys. Rev. B}\ }\textbf {\bibinfo {volume} {74}},\ \bibinfo {pages} {024428}
  (\bibinfo {year} {2006})}\BibitemShut {NoStop}%
\bibitem [{\citenamefont {Nag}\ \emph {et~al.}(2022)\citenamefont {Nag},
  \citenamefont {Rani}, \citenamefont {Singh}, \citenamefont {Venkatesh},
  \citenamefont {Sahni}, \citenamefont {Yadav}, \citenamefont {Jha},
  \citenamefont {Bhattacharyya}, \citenamefont {Babu}, \citenamefont {Suresh}
  \emph {et~al.}}]{nag2022cofevsb}%
  \BibitemOpen
  \bibfield  {author} {\bibinfo {author} {\bibfnamefont {J.}~\bibnamefont
  {Nag}}, \bibinfo {author} {\bibfnamefont {D.}~\bibnamefont {Rani}}, \bibinfo
  {author} {\bibfnamefont {D.}~\bibnamefont {Singh}}, \bibinfo {author}
  {\bibfnamefont {R.}~\bibnamefont {Venkatesh}}, \bibinfo {author}
  {\bibfnamefont {B.}~\bibnamefont {Sahni}}, \bibinfo {author} {\bibfnamefont
  {A.~K.}\ \bibnamefont {Yadav}}, \bibinfo {author} {\bibfnamefont {S.~N.}\
  \bibnamefont {Jha}}, \bibinfo {author} {\bibfnamefont {D.}~\bibnamefont
  {Bhattacharyya}}, \bibinfo {author} {\bibfnamefont {P.~D.}\ \bibnamefont
  {Babu}}, \bibinfo {author} {\bibfnamefont {K.~G.}\ \bibnamefont {Suresh}},
  \emph {et~al.},\ }\href@noop {} {\bibfield  {journal} {\bibinfo  {journal}
  {Phys. Rev. B}\ }\textbf {\bibinfo {volume} {105}},\ \bibinfo {pages}
  {144409} (\bibinfo {year} {2022})}\BibitemShut {NoStop}%
\bibitem [{\citenamefont {Samanta}\ \emph {et~al.}(2020)\citenamefont
  {Samanta}, \citenamefont {Chaudhuri}, \citenamefont {Singh}, \citenamefont
  {Srihari}, \citenamefont {Nigam},\ and\ \citenamefont
  {Bhobe}}]{samanta2020structural}%
  \BibitemOpen
  \bibfield  {author} {\bibinfo {author} {\bibfnamefont {T.}~\bibnamefont
  {Samanta}}, \bibinfo {author} {\bibfnamefont {S.}~\bibnamefont {Chaudhuri}},
  \bibinfo {author} {\bibfnamefont {S.}~\bibnamefont {Singh}}, \bibinfo
  {author} {\bibfnamefont {V.}~\bibnamefont {Srihari}}, \bibinfo {author}
  {\bibfnamefont {A.~K.}\ \bibnamefont {Nigam}},\ and\ \bibinfo {author}
  {\bibfnamefont {P.~A.}\ \bibnamefont {Bhobe}},\ }\href@noop {} {\bibfield
  {journal} {\bibinfo  {journal} {J. Alloys Compd.}\ }\textbf {\bibinfo
  {volume} {819}},\ \bibinfo {pages} {153029} (\bibinfo {year}
  {2020})}\BibitemShut {NoStop}%
\bibitem [{\citenamefont {Gui}\ \emph {et~al.}(2021)\citenamefont {Gui},
  \citenamefont {Feng}, \citenamefont {Cao},\ and\ \citenamefont
  {Cava}}]{gui2021ferromagnetic}%
  \BibitemOpen
  \bibfield  {author} {\bibinfo {author} {\bibfnamefont {X.}~\bibnamefont
  {Gui}}, \bibinfo {author} {\bibfnamefont {E.}~\bibnamefont {Feng}}, \bibinfo
  {author} {\bibfnamefont {H.}~\bibnamefont {Cao}},\ and\ \bibinfo {author}
  {\bibfnamefont {R.~J.}\ \bibnamefont {Cava}},\ }\href@noop {} {\bibfield
  {journal} {\bibinfo  {journal} {J. Am. Chem. Soc.}\ }\textbf {\bibinfo
  {volume} {143}},\ \bibinfo {pages} {14342} (\bibinfo {year}
  {2021})}\BibitemShut {NoStop}%
\bibitem [{\citenamefont {Rossiter}(1991)}]{rossiter1991electrical}%
  \BibitemOpen
  \bibfield  {author} {\bibinfo {author} {\bibfnamefont {P.~L.}\ \bibnamefont
  {Rossiter}},\ }\href@noop {} {\emph {\bibinfo {title} {The electrical
  resistivity of metals and alloys}}},\ Vol.~\bibinfo {volume} {6}\ (\bibinfo
  {publisher} {Cambridge university press},\ \bibinfo {year}
  {1991})\BibitemShut {NoStop}%
\bibitem [{\citenamefont {Gr{\"u}neisen}(1933)}]{gruneisen1933abhangigkeit}%
  \BibitemOpen
  \bibfield  {author} {\bibinfo {author} {\bibfnamefont {E.}~\bibnamefont
  {Gr{\"u}neisen}},\ }\href@noop {} {\bibfield  {journal} {\bibinfo  {journal}
  {Ann. Phys. (Berl.)}\ }\textbf {\bibinfo {volume} {408}},\ \bibinfo {pages}
  {530} (\bibinfo {year} {1933})}\BibitemShut {NoStop}%
\bibitem [{\citenamefont {Bainsla}\ \emph
  {et~al.}(2015{\natexlab{d}})\citenamefont {Bainsla}, \citenamefont {Mallick},
  \citenamefont {Coelho}, \citenamefont {Nigam}, \citenamefont {Varaprasad},
  \citenamefont {Takahashi}, \citenamefont {Alam}, \citenamefont {Suresh},\
  and\ \citenamefont {Hono}}]{bainsla2015high}%
  \BibitemOpen
  \bibfield  {author} {\bibinfo {author} {\bibfnamefont {L.}~\bibnamefont
  {Bainsla}}, \bibinfo {author} {\bibfnamefont {A.}~\bibnamefont {Mallick}},
  \bibinfo {author} {\bibfnamefont {A.~A.}\ \bibnamefont {Coelho}}, \bibinfo
  {author} {\bibfnamefont {A.}~\bibnamefont {Nigam}}, \bibinfo {author}
  {\bibfnamefont {B.~S. D. C.~S.}\ \bibnamefont {Varaprasad}}, \bibinfo
  {author} {\bibfnamefont {Y.~K.}\ \bibnamefont {Takahashi}}, \bibinfo {author}
  {\bibfnamefont {A.}~\bibnamefont {Alam}}, \bibinfo {author} {\bibfnamefont
  {K.~G.}\ \bibnamefont {Suresh}},\ and\ \bibinfo {author} {\bibfnamefont
  {K.}~\bibnamefont {Hono}},\ }\href@noop {} {\bibfield  {journal} {\bibinfo
  {journal} {J. Magn. Magn. Mater.}\ }\textbf {\bibinfo {volume} {394}},\
  \bibinfo {pages} {82} (\bibinfo {year} {2015}{\natexlab{d}})}\BibitemShut
  {NoStop}%
\bibitem [{\citenamefont {Rani}\ \emph {et~al.}(2019)\citenamefont {Rani},
  \citenamefont {Bainsla}, \citenamefont {Suresh},\ and\ \citenamefont
  {Alam}}]{rani2019experimental}%
  \BibitemOpen
  \bibfield  {author} {\bibinfo {author} {\bibfnamefont {D.}~\bibnamefont
  {Rani}}, \bibinfo {author} {\bibfnamefont {L.}~\bibnamefont {Bainsla}},
  \bibinfo {author} {\bibfnamefont {K.~G.}\ \bibnamefont {Suresh}},\ and\
  \bibinfo {author} {\bibfnamefont {A.}~\bibnamefont {Alam}},\ }\href@noop {}
  {\bibfield  {journal} {\bibinfo  {journal} {J. Magn. Magn. Mater.}\ }\textbf
  {\bibinfo {volume} {492}},\ \bibinfo {pages} {165662} (\bibinfo {year}
  {2019})}\BibitemShut {NoStop}%
\bibitem [{\citenamefont {Gupta}\ \emph {et~al.}(2022)\citenamefont {Gupta},
  \citenamefont {Chakraborty}, \citenamefont {Pakhira}, \citenamefont
  {Barreteau}, \citenamefont {Crivello}, \citenamefont {Bandyopadhyay},
  \citenamefont {Greneche}, \citenamefont {Alleno},\ and\ \citenamefont
  {Mazumdar}}]{gupta2022coexisting}%
  \BibitemOpen
  \bibfield  {author} {\bibinfo {author} {\bibfnamefont {S.}~\bibnamefont
  {Gupta}}, \bibinfo {author} {\bibfnamefont {S.}~\bibnamefont {Chakraborty}},
  \bibinfo {author} {\bibfnamefont {S.}~\bibnamefont {Pakhira}}, \bibinfo
  {author} {\bibfnamefont {C.}~\bibnamefont {Barreteau}}, \bibinfo {author}
  {\bibfnamefont {J.-C.}\ \bibnamefont {Crivello}}, \bibinfo {author}
  {\bibfnamefont {B.}~\bibnamefont {Bandyopadhyay}}, \bibinfo {author}
  {\bibfnamefont {J.~M.}\ \bibnamefont {Greneche}}, \bibinfo {author}
  {\bibfnamefont {E.}~\bibnamefont {Alleno}},\ and\ \bibinfo {author}
  {\bibfnamefont {C.}~\bibnamefont {Mazumdar}},\ }\href@noop {} {\bibfield
  {journal} {\bibinfo  {journal} {Physical Review B}\ }\textbf {\bibinfo
  {volume} {106}},\ \bibinfo {pages} {115148} (\bibinfo {year}
  {2022})}\BibitemShut {NoStop}%
\bibitem [{\citenamefont {Feng}\ \emph {et~al.}(2015)\citenamefont {Feng},
  \citenamefont {Chen}, \citenamefont {Yuan}, \citenamefont {Zhou},\ and\
  \citenamefont {Chen}}]{feng2015effect}%
  \BibitemOpen
  \bibfield  {author} {\bibinfo {author} {\bibfnamefont {Y.}~\bibnamefont
  {Feng}}, \bibinfo {author} {\bibfnamefont {H.}~\bibnamefont {Chen}}, \bibinfo
  {author} {\bibfnamefont {H.}~\bibnamefont {Yuan}}, \bibinfo {author}
  {\bibfnamefont {Y.}~\bibnamefont {Zhou}},\ and\ \bibinfo {author}
  {\bibfnamefont {X.}~\bibnamefont {Chen}},\ }\href@noop {} {\bibfield
  {journal} {\bibinfo  {journal} {Journal of Magnetism and Magnetic Materials}\
  }\textbf {\bibinfo {volume} {378}},\ \bibinfo {pages} {7} (\bibinfo {year}
  {2015})}\BibitemShut {NoStop}%
\end{thebibliography}
%

\end{document}